\newcommand\beq{\begin{eqnarray}}
\newcommand\eeq{\end{eqnarray}}
\newcommand\meff{m_{\rm eff}}
\newcommand\missET{E_T^{\rm miss}}

\def\lsim{\mathrel{\rlap{\lower4pt\hbox{$\sim$}}
    \raise1pt\hbox{$<$}}}                
\def\gsim{\mathrel{\rlap{\lower4pt\hbox{$\sim$}}
    \raise1pt\hbox{$>$}}}

\documentclass[
amsmath,
prd,nofootinbib,floatfix,11pt 
]{revtex4}

\allowdisplaybreaks
\usepackage{graphicx}
\usepackage{setspace}

\begin{document}
\renewcommand{\theequation}{\arabic{section}.\arabic{equation}}
\begin{flushright}
ANL-HEP-PR-11-32
\end{flushright}

\title{\Large%
\baselineskip=21pt
Large Hadron Collider reach for supersymmetric models 
with compressed mass spectra}

\author{Thomas J.~LeCompte$^1$ and Stephen P. Martin$^{2,3}$}
\affiliation{
{\it 1) Argonne National Laboratory, Argonne IL 60439, USA} 
\\
{\it 2) Department of Physics, Northern Illinois University, 
DeKalb IL 60115, USA} 
\\
{\it 3) Fermi National Accelerator Laboratory, P.O. Box 500, 
Batavia IL 60510, USA}
}

\begin{abstract}\normalsize \baselineskip=15pt 
Many theoretical and experimental results on the reach of the Large 
Hadron Collider are based on the mSUGRA-inspired scenario with universal 
soft supersymmetry breaking parameters at the apparent gauge 
coupling unification scale. We study signals for 
supersymmetric models in which the sparticle mass range is compressed 
compared to mSUGRA, using cuts like those employed by ATLAS for 2010 data. 
The acceptance and the cross-section times acceptance are found for 
several model lines that employ a compression parameter to smoothly 
interpolate between the mSUGRA case and the extreme case of degenerate 
gaugino masses at the weak scale. For models with moderate compression, 
the reach is not much worse, and can even be substantially better, than 
the mSUGRA case. For very compressed mass spectra, the acceptances are 
drastically reduced, especially when a more stringent effective mass 
cut is chosen.
\end{abstract}


\maketitle

\tableofcontents

\vfill\eject
\baselineskip=16pt

\setcounter{footnote}{1}
\setcounter{page}{2}
\setcounter{figure}{0}
\setcounter{table}{0}

\section{Introduction}
\label{sec:intro}
\setcounter{equation}{0}
\setcounter{footnote}{1}

The continuing explorations of the ATLAS \cite{ATLASTDR} and CMS 
\cite{CMSTDR} experiments at the Large Hadron Collider (LHC) are testing 
the idea that supersymmetry \cite{SUSYreviews} (SUSY) is the solution to 
the hierarchy problem associated with the small ratio of the electroweak 
scale to the Planck scale and other high energy scales. 
Already, there are significant bounds from 
both ATLAS \cite{ATLASL}-\cite{Aad:2011xm} and CMS 
\cite{Khachatryan:2011tk}-\cite{Chatrchyan:2011wb} on certain 
supersymmetric models, especially those formulated in terms of universal 
soft supersymmetry-breaking parameters at the scale of apparent gauge 
coupling unification, the  so-called ``mSUGRA" or ``CMSSM" scenario. 
However, the essential idea of 
supersymmetry is simply that of a symmetry connecting fermion and boson 
degrees of freedom, and the unknown features of the supersymmetry 
breaking mechanism allow for a much more diverse variety of 
possibilities. It is therefore always important to consider the extent to 
which the limits obtained from experimental searches depend on the 
specific model assumptions.

The input parameters for mSUGRA models are a universal gaugino mass 
$M_{1/2}$, a universal scalar mass $m_0$, a common scalar cubic 
coupling parameter $A_0$, the ratio of the Higgs vacuum expectation values 
$\tan\beta = \langle v_u \rangle/\langle v_d \rangle$, and the sign of 
the supersymmetric Higgs mass parameter $\mu$. Existing model-dependent LHC 
searches exploit the fact that supersymmetric particle production and 
decay in mSUGRA-like models typically result in both energetic jets from 
gluino or squark decays and large missing transverse energy $\missET$ due 
to the presence in each superpartner decay chain of a neutral, 
weakly-interacting, stable lightest supersymmetric particle (LSP). 
In mSUGRA models, there is typically a hierarchy of a factor of 6 
or so between the masses of the gluino and the neutralino LSP, leading to robust 
signals. Along with the gluino, the dominant production mechanisms 
involve squarks, which are never much lighter than 
the gluino in viable mSUGRA models.

If the supersymmetric mass spectrum is compressed, that is, if ratio 
of the mass scale of the strongly interacting superpartners 
(the gluino and squarks) and the LSP mass is smaller, then one may
expect that the detection efficiency and acceptance will be lowered,
because there will often be less energetic jets and leptons from 
the decays, as well as less $\missET$. Cuts on the relevant kinematic 
quantities  are necessary for triggering and for reduction of 
backgrounds, so that a sufficiently compressed supersymmetric 
mass spectrum will have greatly reduced discovery potential even if there 
is a large production cross-section.
The purpose of this paper is 
to study this issue quantitatively, using as 
an example selections employed by ATLAS in the study of 35 pb$^{-1}$ of 
data collected in 2010 at 
$\sqrt{s} = 7$ TeV \cite{ATLASL}-\cite{Aad:2011xm}.
Other recent\footnote{The reduced reach of hadron colliders 
for compressed superpartner mass spectra was studied as long ago
as the mid-1980s \cite{Baer:1986ki}, in the context of $< 100$ GeV gluino 
searches by the UA1 detector at the CERN SPS $p\overline p$ collider.}
studies of non-mSUGRA scenarios at the LHC from other viewpoints can be 
found, for example, in \cite{Baer:2007uz}-\cite{Akula:2011dd}.

In this paper we are interested in the consequences 
of a compressed superpartner mass spectrum rather than the 
model-building ideas that might be involved in such a scenario. 
However, it is worth noting that in non-mSUGRA models, it is 
perfectly sensible to choose the three gaugino mass parameters 
to be in any desired proportion at the apparent gauge coupling 
unification scale, which in turn allows any desired ratio of 
gaugino masses at the TeV scale after renormalization group running. 
A huge number of different model building ideas can realize this, 
and one restricted class will be mentioned in subsection 
\ref{subsec:TTmodels} below. Similarly, the squark and slepton 
mass parameters can be chosen arbitrarily, with no problems 
as long as flavor symmetries are respected (as we do in this paper).

The rest of this paper is organized as follows. Section 
\ref{sec:procedures} describes our procedures. Section 
\ref{sec:compressed}
presents results on the acceptance and on the cross-section times 
acceptance, for several classes of 
models defined with a compression parameter 
that can be continuously dialed to vary the ratio of gluino to LSP 
masses. Subsection \ref{subsec:cmodel} gives results for models with 
light squarks and winos, subsection \ref{subsec:HWmodel} 
for similar models  but with the wino mass parameter taken to be 
large enough so that the squarks and gluino do not decay to 
wino-like intermediate states, and subsection \ref{subsec:HSQmodel} 
similarly discusses models that have heavy squarks. All of these are 
complete SUSY models, but 
are defined without reference to dark matter or other indirect 
observable clues on the superpartner spectrum. In section \ref{subsec:TTmodels}, 
we discuss reach for a class of compressed SUSY models that  
have relatively light top squarks and
are specifically motivated by  the dark matter relic abundance 
inferred from WMAP and other experiments. Section \ref{sec:conclusion}
contains some concluding remarks.

\section{Procedures and signal requirements}
\label{sec:procedures}
\setcounter{equation}{0}
\setcounter{footnote}{1}

For this paper, we used {\tt MadGraph/MadEvent 4.4.62} \cite{MGME} to 
generate hard scattering events using CTEQ6L1 \cite{CTEQ}
parton distribution functions, {\tt Pythia 6.422} \cite{Pythia} for 
decays and showering and hadronization, and PGS 4 \cite{PGS} for detector 
simulation. In compressed SUSY models, it is potentially important to 
match correctly (without overcounting) between matrix-element and 
showering/hadronization 
software generation of additional jets. We do this by generating each 
lowest-order process together with the same process with one additional 
jet at the matrix-element level, followed by MLM matching with 
$P_T$-ordered showers with the shower-$K_T$ scheme with $Q_{\rm cut} = 
100$ GeV, as described in \cite{matching} and implemented in the 
MadGraph/MadEvent package. (Including up to two 
extra jets at the matrix-element level is much more time-consuming, and 
we found with some sample testing that even for very compressed 
superpartner mass spectra it did not make a significant difference with our setup.) For 
the detector simulation, we used the default ATLAS-like parameter 
card file provided with the PGS distribution, but with a jet 
cone size of $\Delta R = 0.4$. Cross-sections were normalized to the 
next-to-leading order output of 
{\tt Prospino 2.1} \cite{prospino}. 

To define signals, we follow (a slightly simplified version of) the ATLAS 
cuts A, C, D for multijets+$\missET$ from ref.~\cite{ATLASJ}, and the 
single lepton plus multijets+$\missET$ signal from ref.~\cite{ATLASL}, 
called L here. (We do not attempt to simulate the ref.~\cite{ATLASJ} 
signal region B, which involves the kinematic variable $m_{T2}$. The
B region is intended to be intermediate between the A and D regions.) 
The signal  requirements are as follows. The minimum number of 
jets is $2$ for signal  A and $3$ for signals C, D, and L, each 
with $p_T > 40$ GeV for A, C, D  and $p_T > 30$ GeV for L. 
These jets must have $|\eta| < 2.5$. The $p_T$ 
of the leading jet must exceed 120~GeV for A, C, D, and 60 GeV 
for L.  
The required number of leptons $\ell = (e,\mu)$ is exactly 0 for A, C, D 
and exactly 1 for L. Here leptons are defined to have $|\eta| < 2.4$ 
(2.47) for muons (electrons), and $p_T^\ell > 10$ GeV in signals A, C, D 
and $p_T^\ell > 20$ GeV in signal L, and to be farther than $\Delta R = 
\sqrt{(\Delta \eta)^2 + (\Delta \phi)^2} > 0.4$ from the nearest jet 
candidate with $|\eta|< 4.9$ and $p_T > 20$ GeV. The effective mass 
$\meff$ is defined as the scalar sum of the $\missET$ and the $p_T$'s of: 
the leading 2 jets for A; the leading 3 jets for C, D; and the leading 
three jets and the leading lepton for L. The A, C, D, L signals require 
$\meff > 500$, 500, 1000, 500 GeV respectively. The missing transverse energy 
$\missET$ must exceed a fraction of $\meff$ in each event, given by $0.3$ 
for A and $0.25$ for C, D, L. For the A, C, D signals, the jets with $p_T 
> 40$ GeV, up to a maximum of 3, are required to be isolated from the 
missing transverse momentum according to $\Delta 
\phi(\vec{p}_T^{\phantom{.}\rm miss},j) > 0.4$. A similar but weaker 
requirement applies to L: $\Delta \phi(\vec{p}_T^{\phantom{.}\rm 
miss},j)> 0.2$ for the three highest-$p_T$ jets. For signal L only, the 
transverse mass $m_T = \sqrt{ 2 (p_T^\ell \missET - \vec{p}_T^{\phantom{.}\ell} 
\cdot \vec{p}_T^{\phantom{.}{\rm miss}}) }$ is required to exceed 100 GeV 
for the single lepton $\ell$.

These signal requirements are summarized in Table \ref{tab:cuts}. Note 
that these cuts automatically imply a lower limit on 
$\missET$ of 150, 125, 250, and 125 GeV for signals A, C, D, and L, 
respectively.
\begin{table}
\begin{tabular}[c]{lcccc}
& A & C & D & L 
\\
\hline
\hline
Number of jets & $\geq 2$ & $\geq 3$ & $\geq 3$ & $\geq 3$ 
\\ 
Leading jet $p_T$ [GeV] & $>120$ & $>120$ & $>120$ & $>60$
\\
Other jet(s) $p_T$ [GeV] & $>40$ & $>40$ & $>40$ & $>30$
\\
$\Delta \phi(\vec{p}_T^{\phantom{.}\rm miss}, \> j_{1,2,3})$
& $>0.4$ & $>0.4$ & $>0.4$ & $>0.2$
\\
$\meff$ [GeV] & \phantom{x}$>500$\phantom{x} & 
\phantom{x}$>500$\phantom{x} & \phantom{x}$>1000$\phantom{x} & 
\phantom{x}$>500$\phantom{x} \\
$\missET/\meff$ & $>0.3$ & $>0.25$ & $>0.25$ & $>0.25$
\\
Number of leptons & $=0$ & $=0$ & $=0$ & $=1$ 
\\
Lepton $p_T$ [GeV] & -- & -- & -- & $>20$
\\
$m_T$ [GeV] & -- & -- & -- & $>100$
\\
\hline
ATLAS $\sigma\times{\rm Acc}$ [pb]  \phantom{xx}& $<1.3$ & $<1.1$ & $<0.11$ & $<0.138$ 
\\
\end{tabular}
\caption{\label{tab:cuts} Summary of cuts for the signals A, C, D, L 
simulated 
here, following ATLAS 2010 data analyses \cite{ATLASL,ATLASJ}.
Also shown on the last line are the ATLAS 95\% CL bounds from 2010 data 
(35 pb$^{-1}$ at $\sqrt{s} = 7$ TeV) on the 
non-Standard 
Model contribution to the cross-section times acceptance in the four 
signal regions.} 
\end{table}

The ATLAS searches using the 2010 data (35 pb$^{-1}$ at $\sqrt{s} = 7$ 
TeV) in \cite{ATLASL,ATLASJ} were given in terms of a grid of mSUGRA 
models with $\tan\beta=3$ and $A_0=0$ (which in the mSUGRA framework are 
ruled out already based on the LEP searches for a scalar Higgs boson). A 
realistic estimate of the relevant backgrounds would require a dedicated 
analysis taking into account specific detailed features of the LHC 
detectors, and will not be attempted here. We will therefore present 
results only for the SUSY signals after the cuts listed above. However, 
ATLAS has also presented limits on non-Standard-Model contributions to 
the $\sigma \times$acceptance as $<$ (1.3, 1.1, 0.11, and 0.138) pb for 
signals (A,C,D,L) respectively. These can be considered 
model-independent, with the caveat that supersymmetry, if 
present, could contribute to the control regions used to estimate 
backgrounds from data in different ways depending on the superpartner 
masses and decays. Also, it is 
very likely that future searches will use modified signal requirements in 
order to extend the reach.

Since our tools for generating events and simulating detector response 
are not the same as those used by ATLAS, the cross-section and acceptance results 
found below clearly cannot be interpreted in exact correspondence to the 
ATLAS ones. However, we have checked that the results of our analysis 
methods correlate well to those in refs.~\cite{ATLASL,ATLASJ} for the 
grid of mSUGRA models used there. For the all-jets signals A,C,D, we find 
cross-section times acceptances that typically agree with the ATLAS 
results (as given in Figures 17a,c,d of the web site referred to by \cite{ATLASJ}) to 
20\% or better, with some fluctuations that appear to be statistical in 
nature. For the 1-lepton signal L, our results for the acceptance are 
similar but tend to be systematically higher than those found in Figures 
6a,b of the web site referred to by \cite{ATLASL}, by typically 20\% to 50\%, again 
with significant statistical fluctuations. Keeping these in mind, at 
least an approximate estimate of the true detector response may be 
gleaned from the results below, and the general trends should be robust.

\section{Results for several classes of compressed SUSY models}
\label{sec:compressed}
\setcounter{equation}{0}
\setcounter{footnote}{1}

\subsection{Models with light squarks and winos}
\label{subsec:cmodel}

\begin{figure}[!tbp]
\begin{minipage}[]{8.3cm}
\includegraphics[width=8.3cm,angle=0]{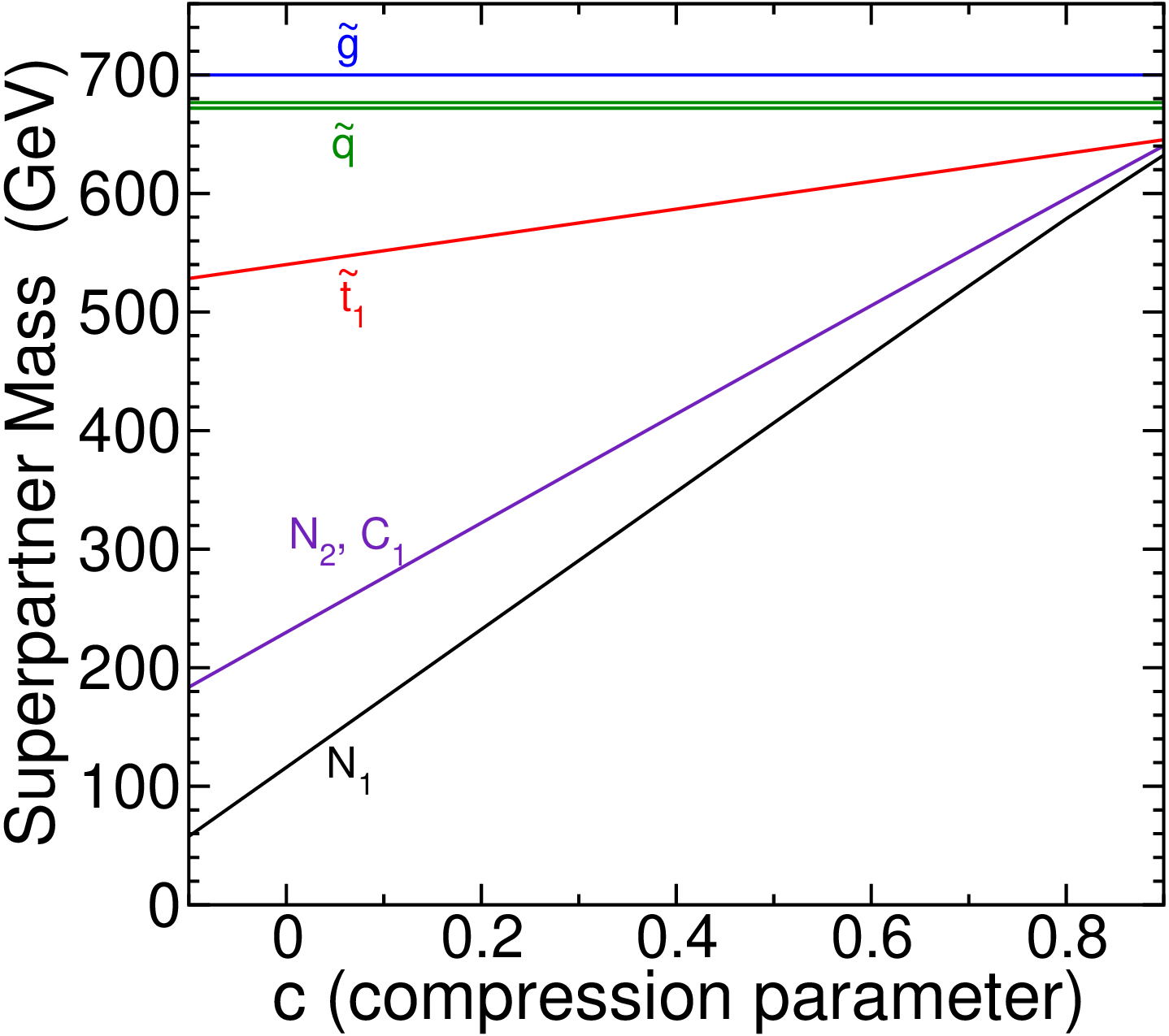}
\end{minipage}
\hspace{0.6cm}
\begin{minipage}[]{4.4cm}
\caption{\label{fig:cmasses}
The masses of the most relevant superpartners for the class of models
defined in subsection \ref{subsec:cmodel}, as a function of the 
compression 
parameter $c$, for fixed $M_{\tilde g} = 700$ GeV. The case $c=0$ 
corresponds to an mSUGRA-like model.}
\end{minipage}
\end{figure}

In this section, we consider a model framework featuring a 
quantity $c$ that parameterizes the compression of the supersymmetric 
mass spectrum. Specifically, we parameterize the electroweak gaugino mass 
parameters at the TeV scale in terms of the gluino physical mass as:
\beq
M_1 \>=\> \left (\frac{1 + 5 c}{6}\right )M_{\tilde g},
\qquad\qquad
M_2 \>=\> \left (\frac{1 + 2 c}{3}\right )M_{\tilde g}.
\label{eq:cmodelgauginomasses}
\eeq
The value $c=0$ gives an 
mSUGRA-like mass spectrum with gaugino masses equal at $M_{\rm 
GUT} = 2.5 \times 10^{16}$ GeV, and $c = 1$ gives a completely compressed 
spectrum in 
which the gluino, wino, and bino masses are equal at the TeV  
scale. The gluino mass $M_{\tilde g}$ is treated as a variable input 
parameter.
We also select $\tan\beta = 10$ and positive $\mu = M_{\tilde g} + 200$ GeV 
to compute the physical masses of charginos $\tilde C_i$ and neutralinos $\tilde N_i$.
The first- and second-family squark masses are:
\beq
m_{\tilde u_R} = m_{\tilde d_R} = m_{\tilde u_L} = 0.96 M_{\tilde g},
\qquad
m^2_{\tilde d_L} = m_{\tilde u_L}^2 - \cos(2\beta) m_W^2,
\eeq
and sleptons are degenerate with the squarks (so too heavy to appear in 
chargino and neutralino two-body decays). The top squark masses are
taken to be $m_{\tilde t_1} = M_{\tilde g} - 160 + c (180 - 0.09 
M_{\tilde g})$ and $m_{\tilde t_2} = M_{\tilde g} + 25$, in GeV.
The lightest Higgs mass is fixed at $m_{h^0} = 115$ GeV, and the heavier 
Higgs masses with $m_{A^0} = 0.96 M_{\tilde g}$.
These values are engineered to provide relatively smoothly varying 
branching ratios as the compression parameter $c$ is varied, although
transitions of $\tilde N_2$ and $\tilde C_1$ decays from on-shell 
to off-shell weak bosons are 
inevitable as the compression increases. 
The choices 
for $\tan\beta$ and $\mu$ are essentially arbitrary,
and not very much would change if they were modified (within
some reasonable range). The reason for the choice for the
parameterization of the stop masses was just to avoid suddenly
turning on or off any 2-body decay modes as the parameter c is varied
within each model line, notably by making sure that the the
gluinos cannot decay to stops by kinematics along the entire
model line.
The masses of the most 
relevant superpartners are shown in Figure \ref{fig:cmasses} for the case 
$M_{\tilde g} = 700$ GeV, 
illustrating the effect of the compression parameter $c$ on the spectrum.
An orthogonal direction in parameter space is obtained by
varying $M_{\tilde g}$, which moves the entire mass spectrum up or 
down for fixed $c$.

\begin{figure}[!tbp]
\mbox{\includegraphics[width=8.3cm,angle=0]{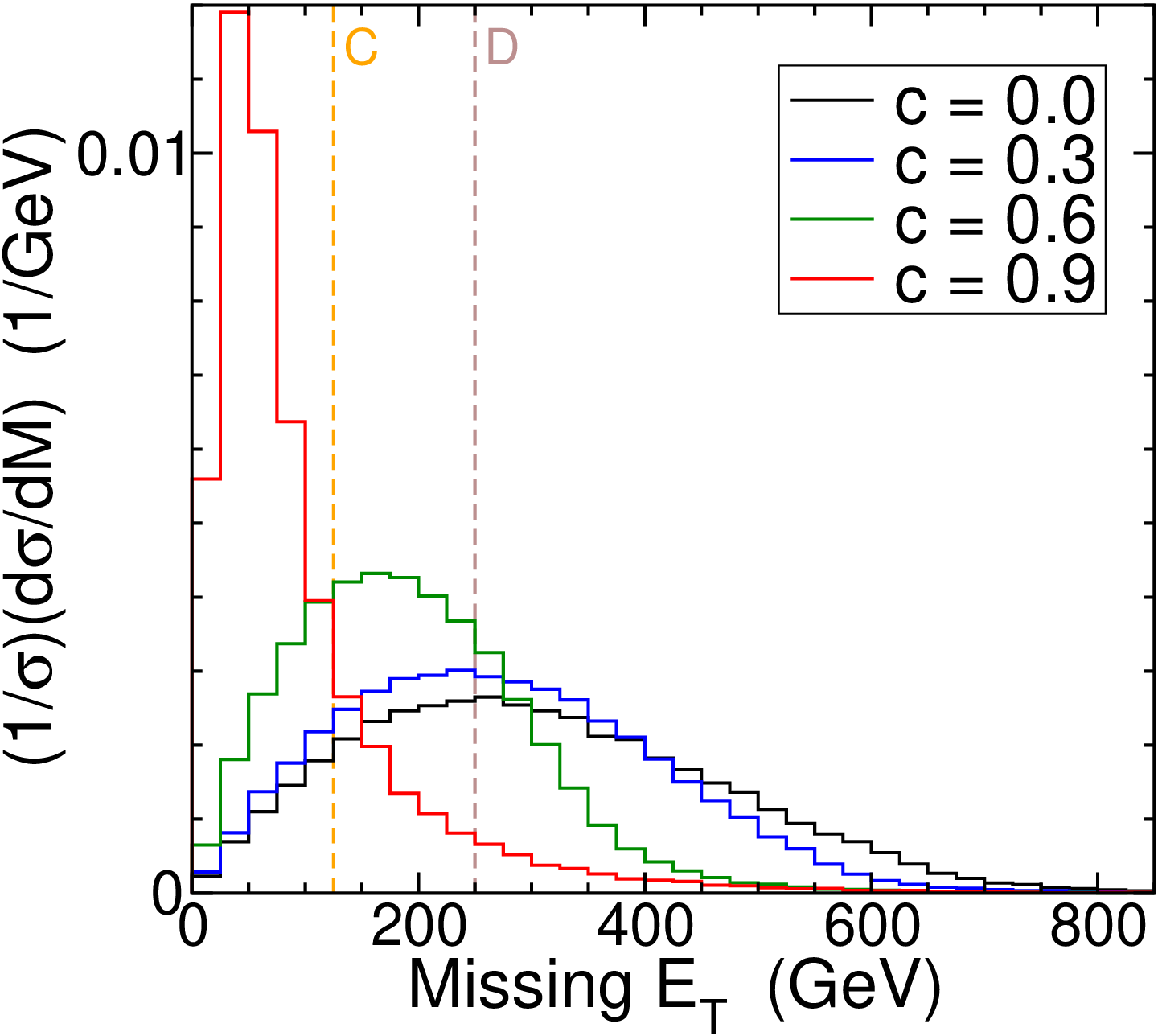}
\includegraphics[width=8.3cm,angle=0]{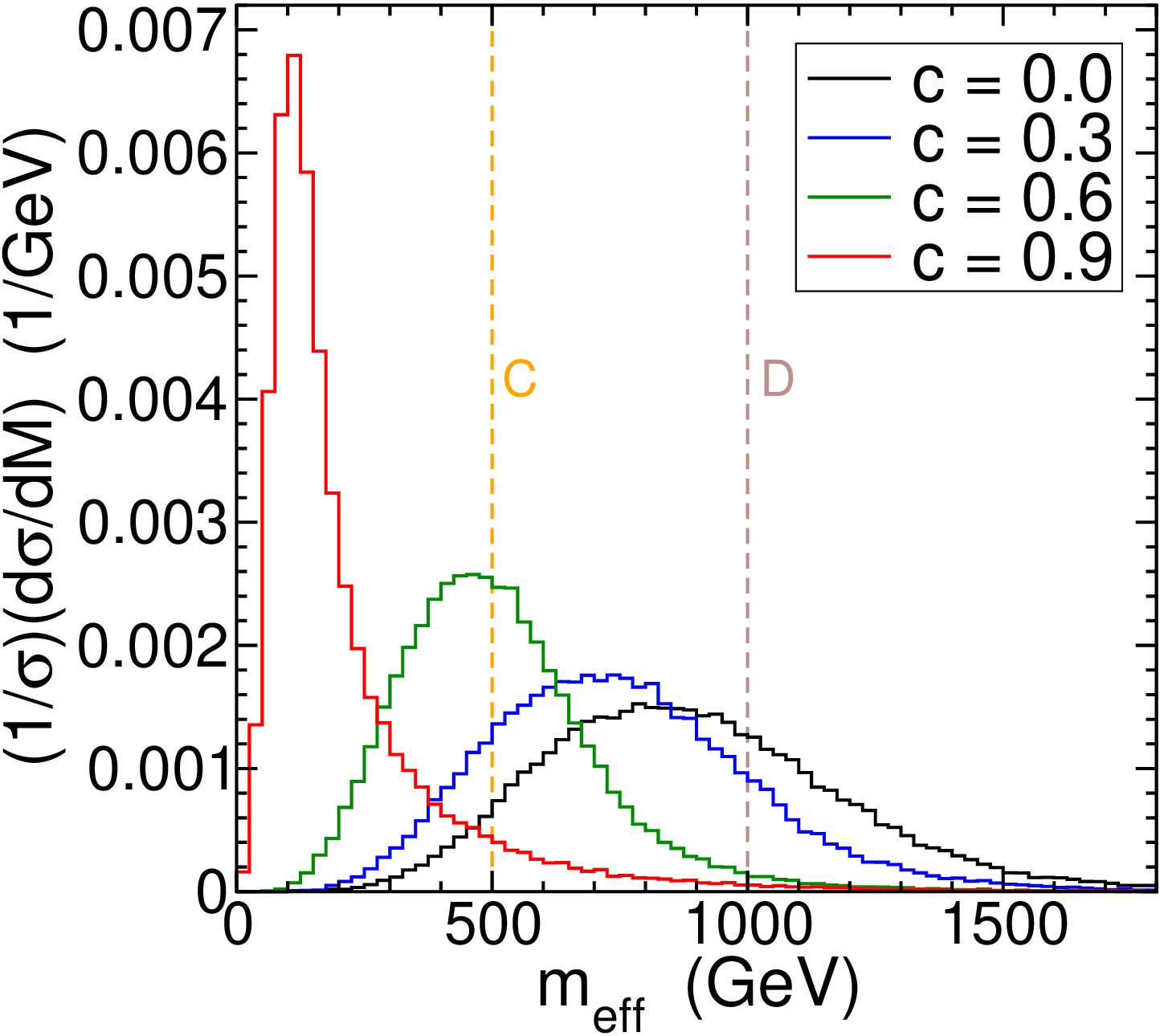}}
\caption{\label{fig:sampledists}
The distributions before cuts of $\missET$ (left panel) and $\meff$ with 3 jets 
included (right panel) for
models described in subsection \ref{subsec:cmodel} with $M_{\tilde g} = 
700$ GeV and $c = 0.0$, 0.3, 0.6, and 0.9, from right to left. The cuts 
for signals C and D are also shown. The $\meff$ distribution decreases more quickly than 
$\missET$ does as $c$ increases.}
\end{figure}

In these models, gluino and squark production dominate at the LHC. 
The gluino decays mostly by the 
two-body mode $\tilde g \rightarrow \overline q \tilde q$ or $q\tilde{\overline q}$, 
and right-handed squarks
decay mostly directly to the LSP, $\tilde q_R \rightarrow q \tilde N_1$,
while left-handed squarks decay mostly to wino-like charginos and 
neutralinos, $\tilde q_L \rightarrow q' \tilde C_1$ and $q \tilde N_2$.
The latter decay through on-shell or off-shell weak bosons:
$\tilde C_1 \rightarrow W^{(*)} \tilde N_1$ and
$\tilde N_2 \rightarrow Z^{(*)} \tilde N_1$, or
$\tilde N_2 \rightarrow h \tilde N_1$ when it is kinematically allowed.
The visible energy in each event from these decays clearly 
decreases as the compression factor $c$ increases, because of the 
reduction in available kinematic phase space. 
To illustrate the effect of this, we show in Figure \ref{fig:sampledists}
the $\missET$ and $\meff$ distributions for $c = 0.0$, 0.3, 0.6, and 0.9 in the case $M_{\tilde g} = 700$ GeV. 
The softening of these distributions becomes
drastic as $c$ approaches 1, leading to a 
more difficult search, at least 
by the usual methods.

\begin{figure}[!tbp]
\mbox{\includegraphics[width=8.3cm,angle=0]{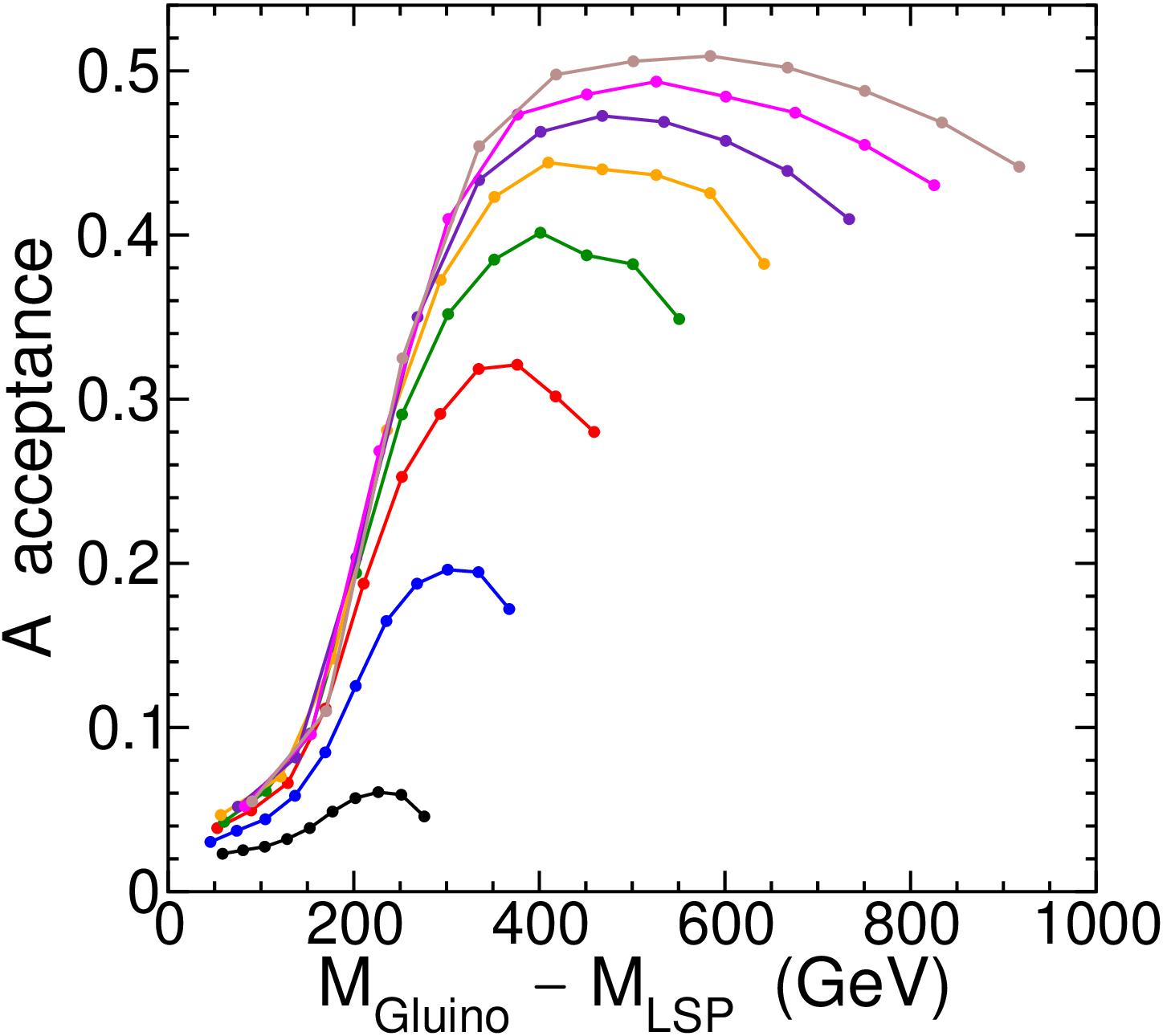}
\includegraphics[width=8.3cm,angle=0]{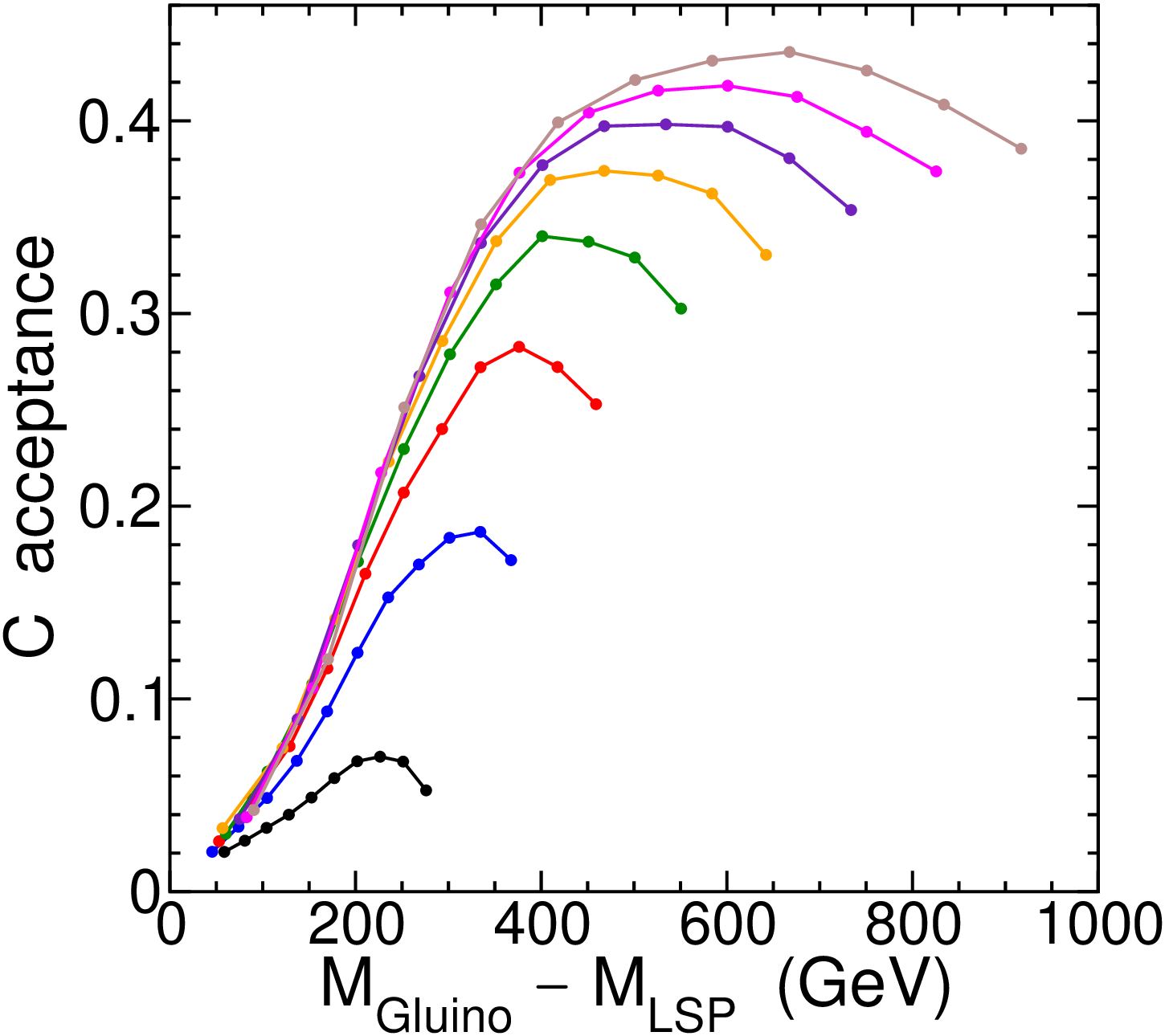}}

\vspace{0.1in}

\mbox{\includegraphics[width=8.3cm,angle=0]{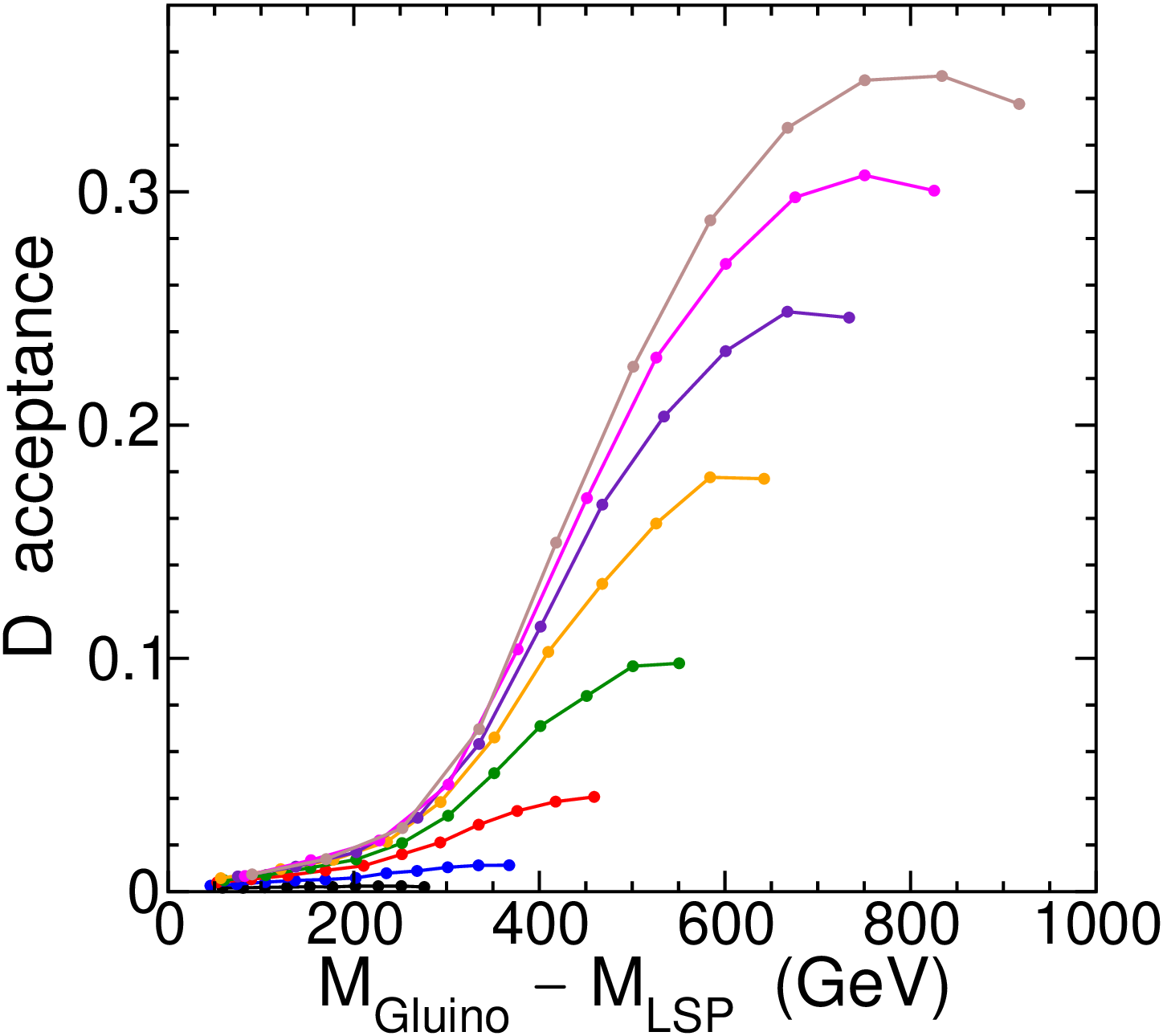}
\includegraphics[width=8.3cm,angle=0]{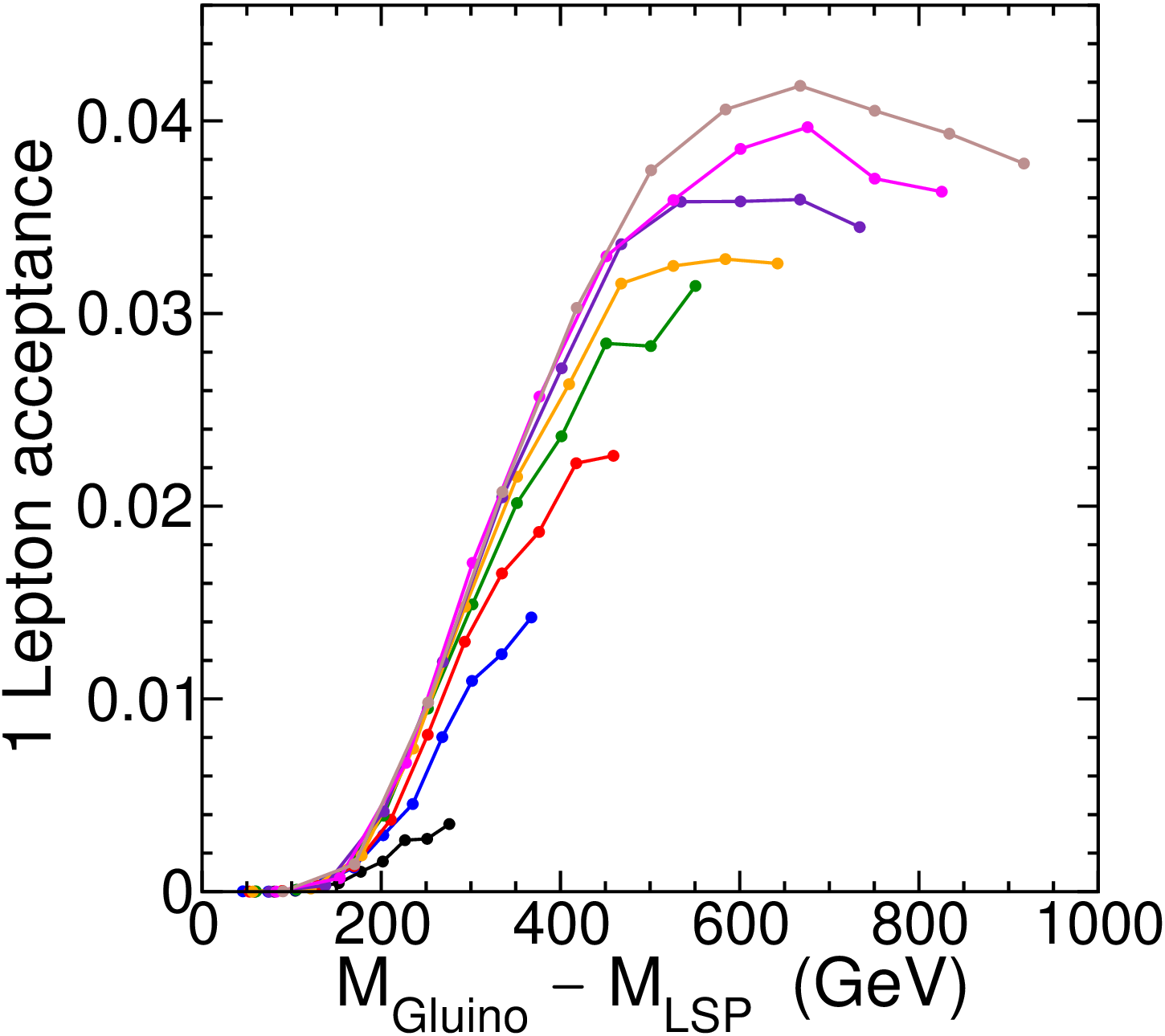}}
\caption{\label{fig:acc}
The acceptances for model lines defined in section \ref{subsec:cmodel} 
as a function of $M_{\tilde g} - M_{{\tilde N_1}}$, obtained by 
varying the gaugino mass compression factor $c$. 
The lines from bottom to top correspond to $M_{\tilde g} = 300, 400, 
\ldots, 1000$ GeV.
The dots on each line correspond to, from right to left,
$c = -0.1, 0, 0.1, \ldots 0.9$, with $c=0$ corresponding to the 
mSUGRA-like case and $c=1$ to a completely compressed gaugino spectrum. 
The four panels are for the four sets of cuts A, C, D, and L.}
\end{figure}

Figure \ref{fig:acc} shows the acceptances
for signals A, C, D, L for $M_{\tilde g} = 300, 400, 
\ldots, 1000$ GeV, with the compression factor varying in the range
$-0.1 < c < 0.9$, as a function of the gluino-LSP mass difference
$M_{\tilde g} - M_{{\tilde N_1}}$.
The acceptances for all four signals become sharply reduced when
the gluino-LSP mass difference decreases below 200 GeV for the A and C
signals, with an even stronger reduction for signals D and L.
The single cut most responsible for decreasing the signal in each case is 
the requirement of a minimum $m_{\rm eff}$.

An interesting feature seen in Figure \ref{fig:acc} is that for fixed 
$M_{\tilde g}$, the acceptances often actually increase with $c$ for low 
$c$, especially when $M_{\tilde g}$ is large and especially for signals A 
and C. For a fixed gluino mass, this leads to a maximum acceptance for 
models that are somewhat more compressed than mSUGRA, which may seem 
counterintuitive. The interplay between $c$ and the event kinematics is 
complicated for cascade decays. The relevant cut is the one on the ratio 
$\missET/\meff$; the $\meff$ distribution rapidly becomes softer 
for increasing $c$, while the $\missET$ distribution decreases much more slowly (and can even increase), allowing more events to pass the ratio cut. 
This effect can be discerned in Figure \ref{fig:sampledists}, for example.
(A similar effect was noted in \cite{Alves:2010za}.)
The events 
responsible for this effect are those with multi-stage decays of left-handed 
squarks through 
the wino-like states $\tilde C_1$ and $\tilde N_2$, and 
the largest effect comes specifically from those events in which both 
winos move in the same direction and both LSPs in the opposite direction 
relative to a fixed direction perpendicular to the beam. For the component of the signal 
coming from left-handed squark production, the
$\missET$ distribution 
actually becomes significantly harder for increasing $c$, 
especially for heavier superpartners. In contrast, for 
events with single-stage decays $\tilde q_R \rightarrow q \tilde N_1$, 
the acceptance decreases monotonically with increasing compression factor 
$c$.

\begin{figure}[!tbp]
\mbox{\includegraphics[width=8.3cm,angle=0]{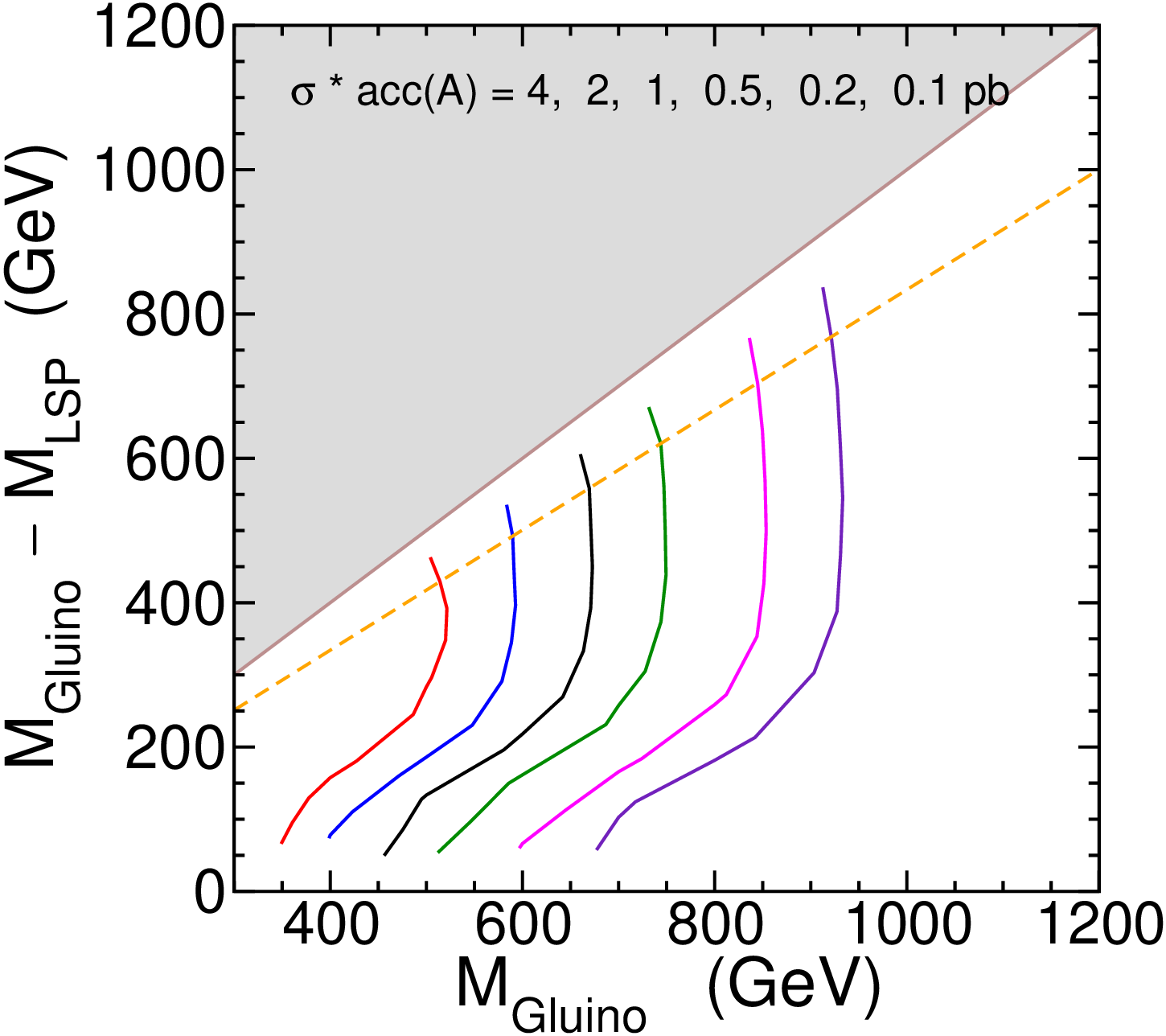}
\includegraphics[width=8.3cm,angle=0]{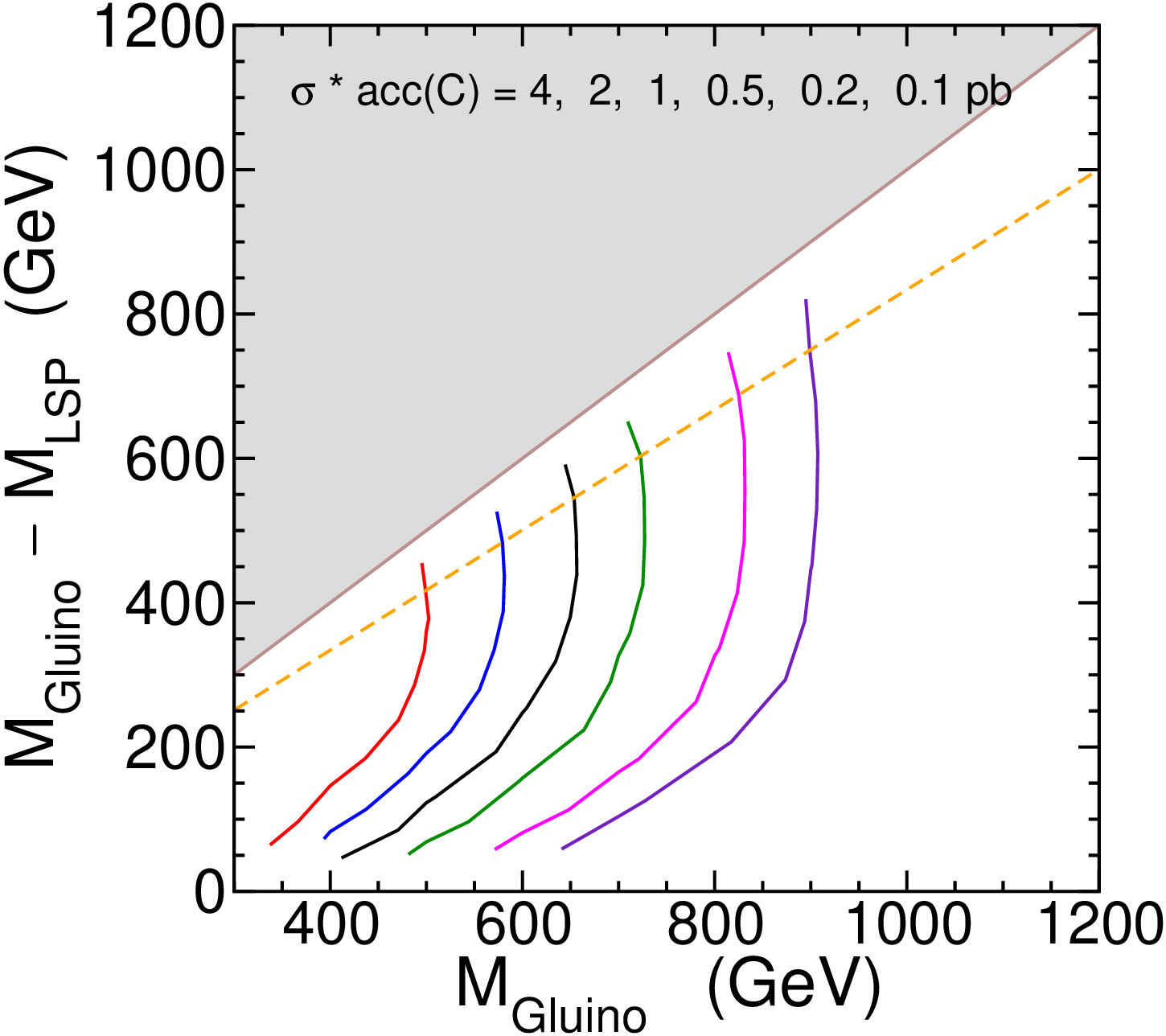}}

\vspace{0.2in}

\mbox{\includegraphics[width=8.3cm,angle=0]{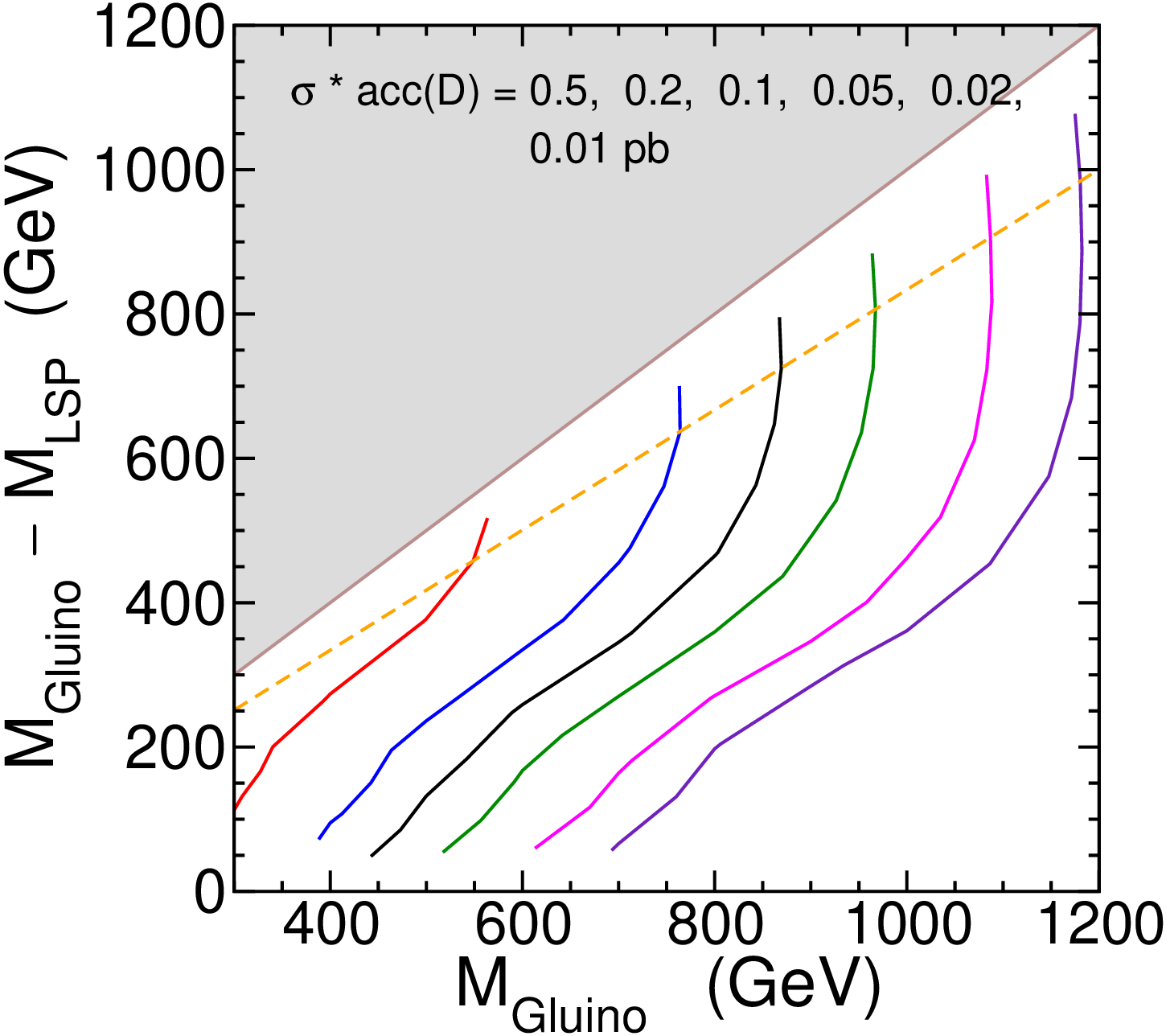}
\includegraphics[width=8.3cm,angle=0]{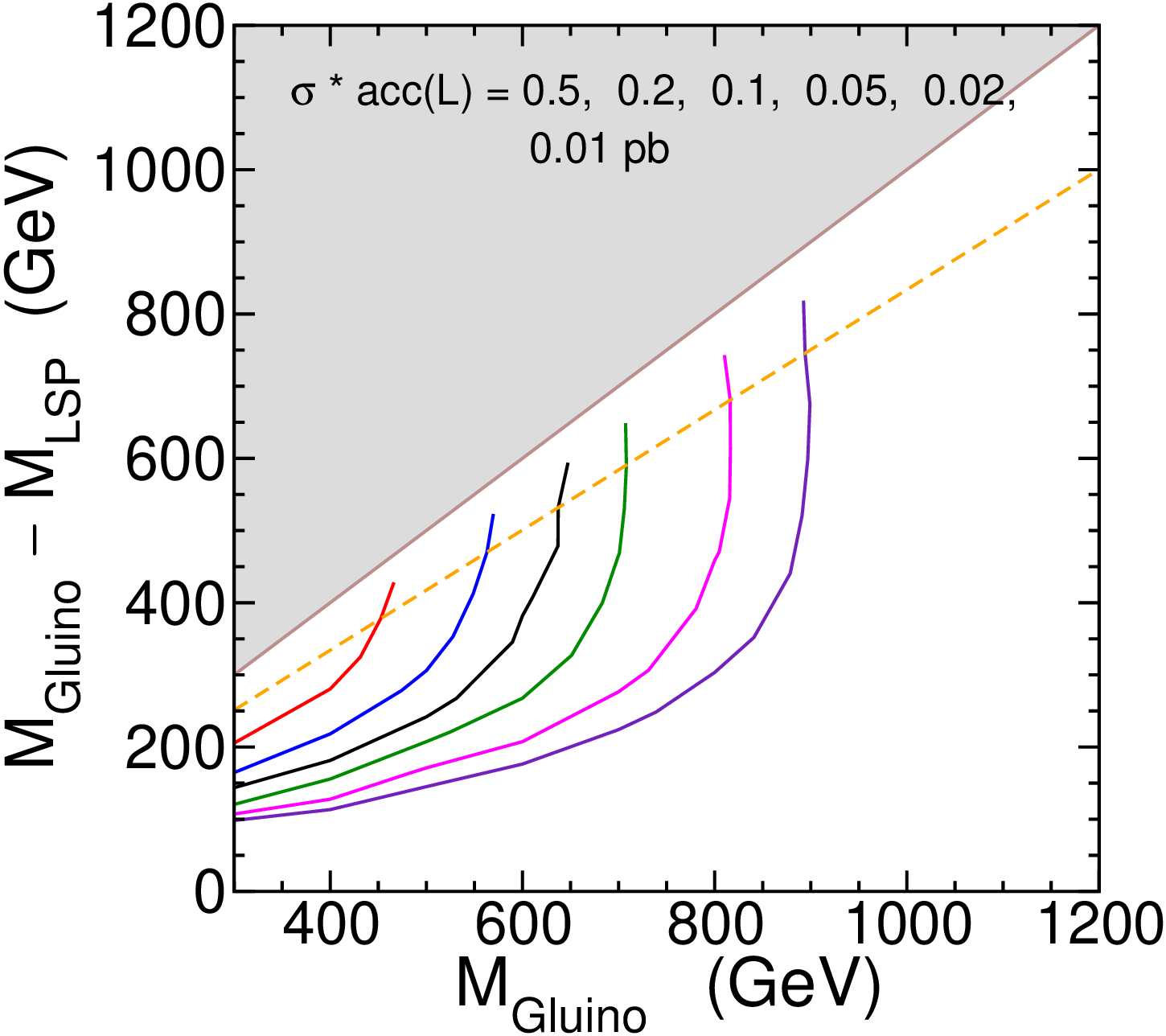}}
\caption{\label{fig:sigacc}
Contours of cross-section times acceptance for the models defined in section 
\ref{subsec:cmodel}, 
in the $M_{\tilde g} - M_{{\tilde N_1}}$ vs. $M_{\tilde g}$ plane, obtained by varying the gaugino mass compression parameter $c$ between $-0.1$ and $0.9$. 
The dashed line corresponds to the mSUGRA-like case $c=0$, with 
increased compression lower in the plane.
The four panels correspond to cuts 
A, C, D, and L.}
\end{figure}

Figure \ref{fig:sigacc} shows contours of the total cross section times 
acceptances for the same models, in the plane of $M_{\tilde g}$ and 
$M_{\tilde g} - M_{\tilde N_1}$. The dashed line indicates the 
mSUGRA-like 
case of $c=0$, and larger compression factor $c$ occurs lower in each 
plot. The shaded region corresponds to the case where the LSP is not a 
neutralino. From this, one infers for example that (to the extent that 
background levels can be established independent of the signal), if one 
can set a limit of 850 GeV on $M_{\tilde g}$ for mSUGRA-like models in 
this class with a given set of data using signal D, then the limit would 
be less than 650 (500) GeV if $M_{\tilde g} - M_{\tilde N_1}$ is as small 
as 300 (100) GeV. As a caveat, one might suspect that the more compressed models 
could also lead to a higher signal contamination of the background 
control regions, so that the real-world limit would be even weaker, but 
establishing this effect would require a dedicated background analysis that is 
beyond the scope of this paper.
\vspace{0.2in}

\clearpage

\subsection{Models with heavy winos}
\label{subsec:HWmodel}
\setcounter{footnote}{1}

In the models of the previous subsection, the wino-like chargino and 
neutralino $\tilde C_1$ and $\tilde N_2$ played an important role in the 
cascade decays of left-handed squarks. In this section, we consider a 
variation on this class of models in which the wino-like states 
decouple from LHC phenomenology because they are 
heavier than the squarks and the gluino. This is never a feature of 
mSUGRA models, but it is actually motivated 
\cite{KaneKing,compressedSUSYa,compressedSUSYc}
as a solution to the 
supersymmetric little hierarchy problem. The essential reason for this is 
that in models with a larger ratio $M_2/M_3$ than in mSUGRA, 
the renormalization
group evolution provides for Higgs potential parameters that require much less
tuning to obtain the observed electroweak breaking scale.
We therefore consider models just like the ones of the previous section, 
but with
\beq
M_1 &=& \left (\frac{1 + 5 c}{6}\right )M_{\tilde g},
\\
M_2 &=& M_{\tilde g} + 100\>\,{\rm GeV}
\eeq
replacing eq.~(\ref{eq:cmodelgauginomasses}) at the TeV scale. Thus the gluino-LSP 
compression is still parameterized by $c$ in the same way as before, but 
the winos are heavy, so that the most important superpartner masses are 
just as in Figure \ref{fig:cmasses} with $\tilde N_2$ and $\tilde C_1$ 
removed. As a result, all first- and second-family squarks now 
decay 
directly to the LSP: $\tilde q \rightarrow q \tilde N_1$. The gluino has 
direct two-body decays
to quarks and squarks as before. This means that 
signals relying on isolated leptons are absent.\footnote{In variations 
on this type of model with larger $M_{\tilde g} - M_{\tilde t_1}$, 
in which decays $\tilde g \rightarrow t 
\tilde t_1$ 
dominate, leptonic signatures from the top decays are important.
See subsection \ref{subsec:TTmodels} for an example of this alternative.}

\begin{figure}[!tbp]
\mbox{\includegraphics[width=8.3cm,angle=0]{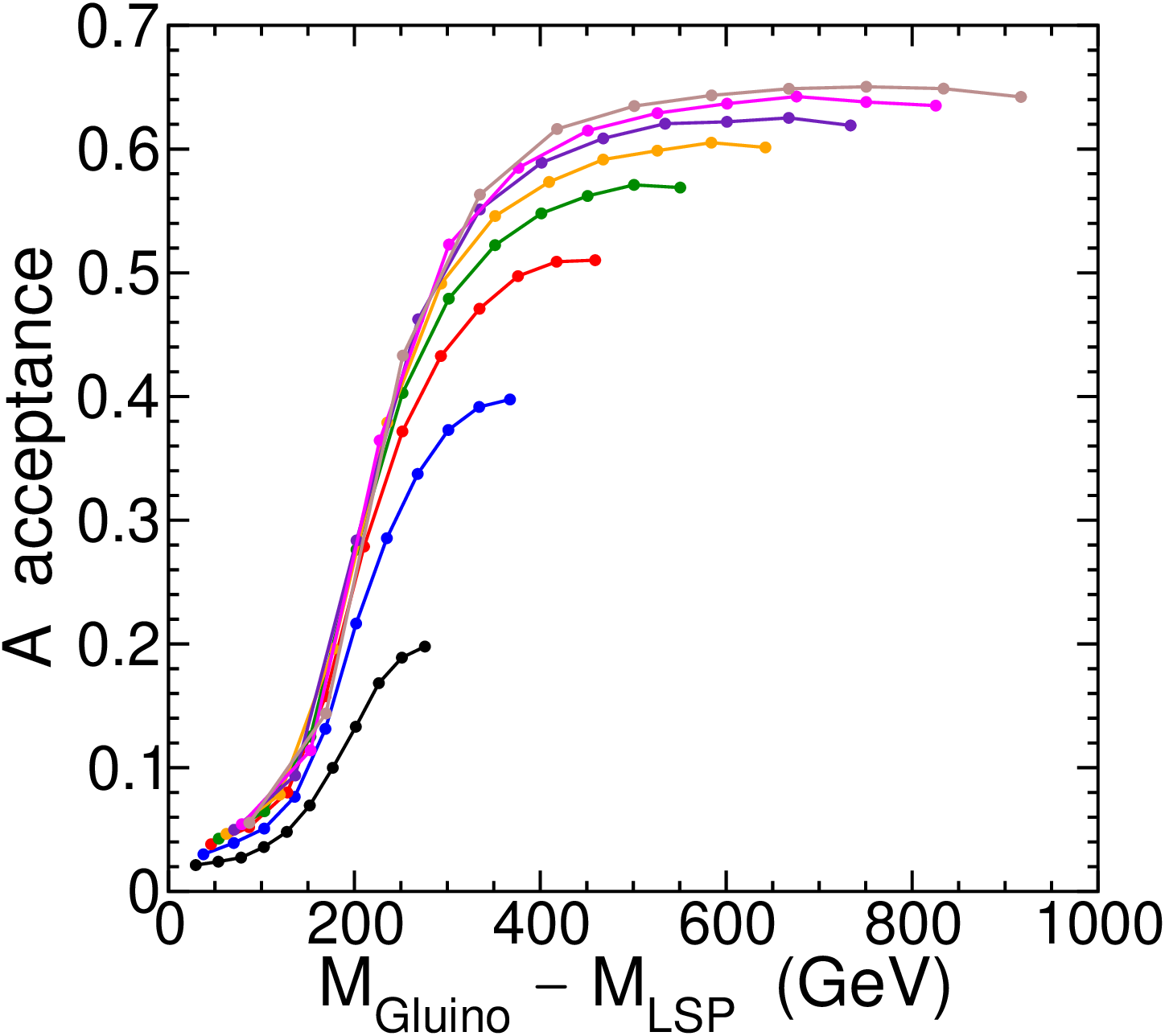}
\includegraphics[width=8.3cm,angle=0]{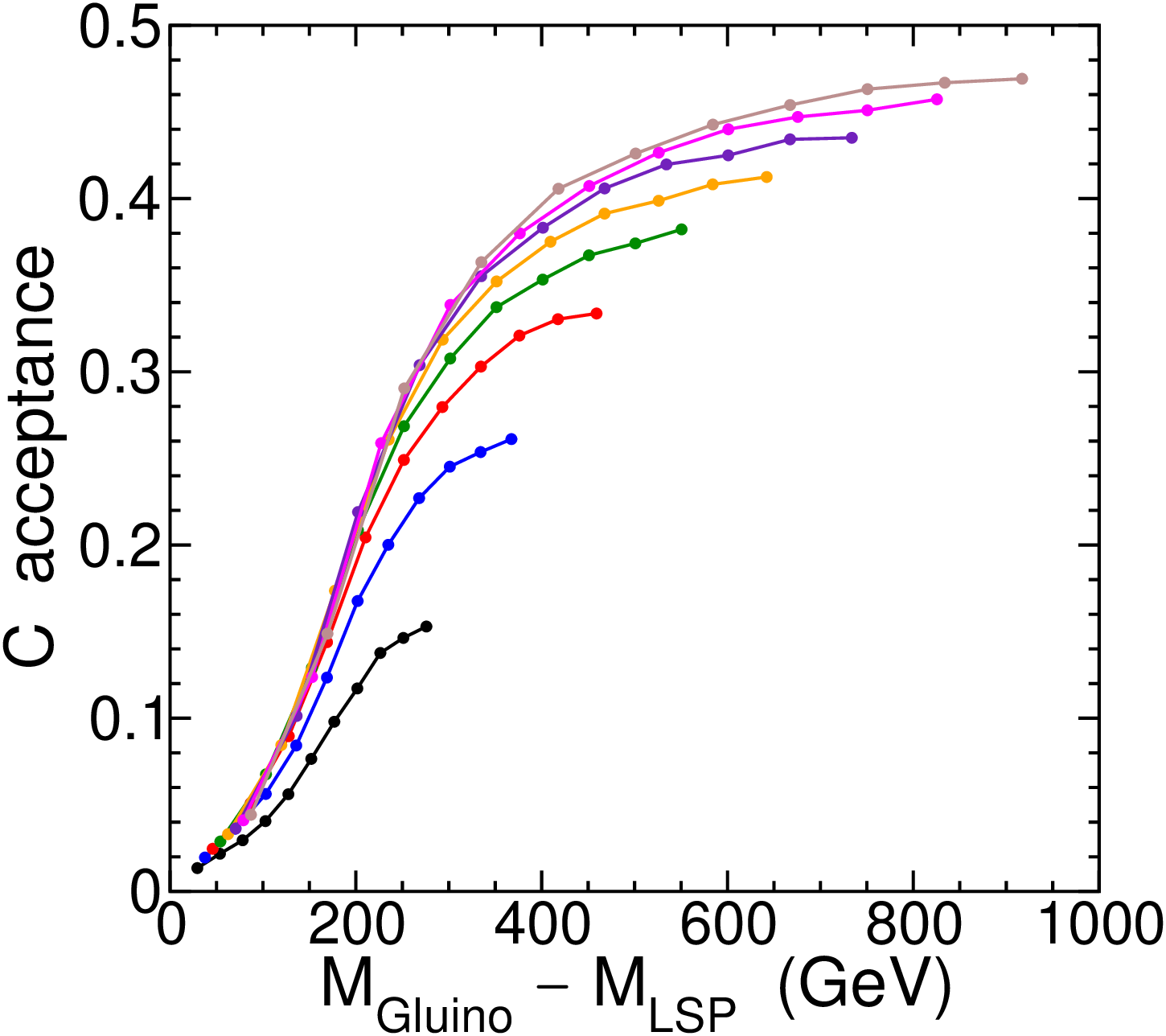}}

\vspace{0.1in}

\mbox{
\begin{minipage}[]{8.3cm}
\includegraphics[width=8.3cm,angle=0]{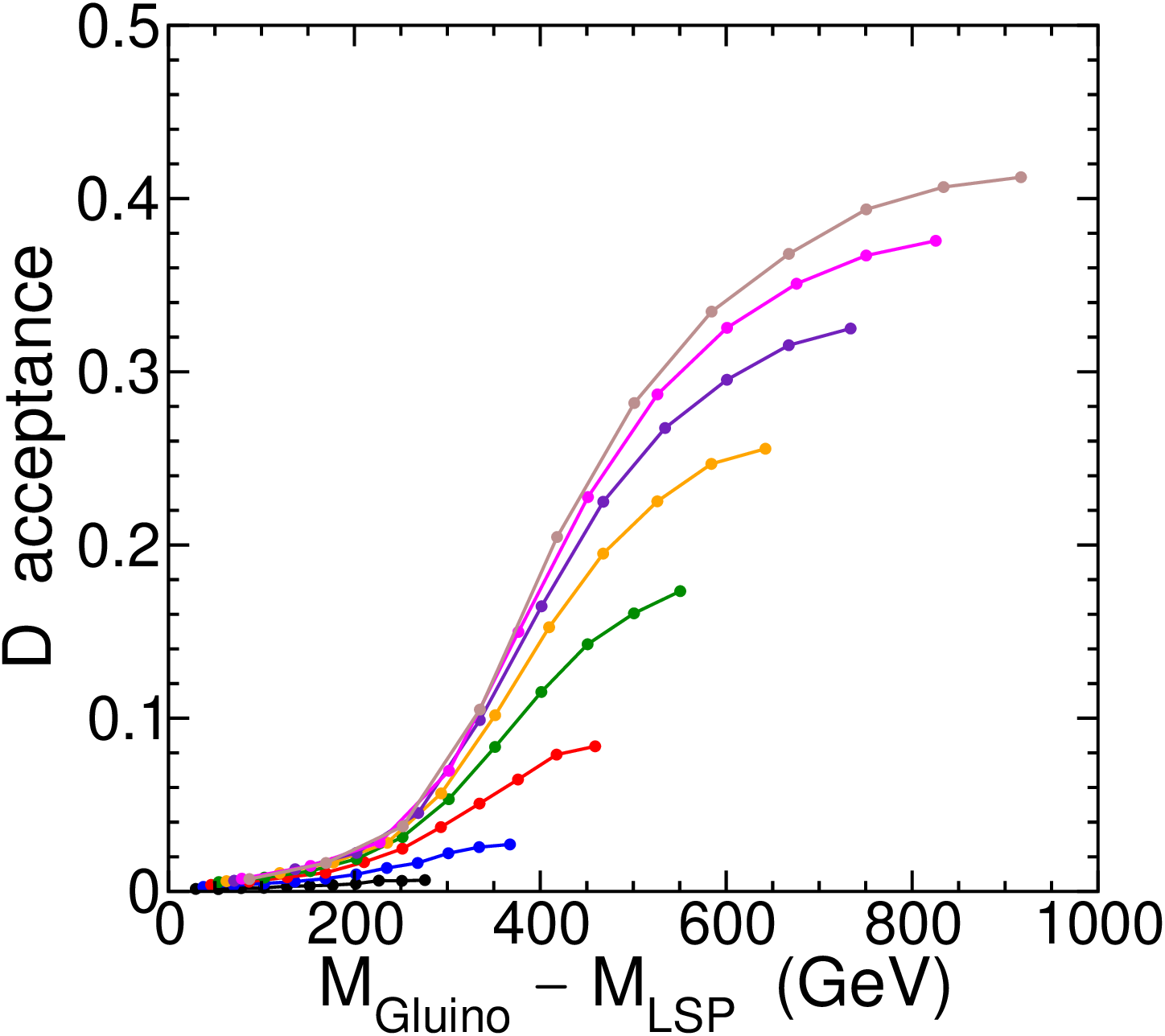}
\end{minipage}
\hspace{0.6cm}
\begin{minipage}[]{7.4cm}
\caption{\label{fig:HWacc}
The acceptances for heavy-wino 
model lines defined in section \ref{subsec:HWmodel} with 
$M_{\tilde g} = 300, 400, \ldots, 1000$ GeV, 
as a function of $M_{\tilde g} - M_{{\tilde N_1}}$, obtained by 
varying the compression factor $c$. 
The dots on each line correspond to, from right to left,
$c = -0.1, 0, 0.1, \ldots 0.9$, with $c=0$ 
corresponding to the mSUGRA-like case
and $c=1$ to a degenerate gluino and bino. The three panels are 
for the cuts 
A, C, and D.}
\end{minipage}
\phantom{xxx}
}
\end{figure}

The resulting acceptances for the all-jets signals A, C, D 
for these heavy-wino 
models are shown
in Figure \ref{fig:HWacc} 
for $M_{\tilde g} = 300, 400, 
\ldots, 1000$ GeV, with the compression factor again varying in the range
$-0.1 \leq c \leq 0.9$, as a function of the mass difference
$M_{\tilde g} - M_{{\tilde N_1}}$.
The acceptance for signal L is found to be always extremely small 
for these models, and so is not 
shown.
Unlike the models in the previous subsection, the acceptance is largest 
for low compression and decreases (essentially) monotonically for 
increasing $c$. This is because of the absence of cascade decays here.
At higher compression, the results are qualitatively 
similar to the light-wino models in the previous subsection, with the 
main difference being somewhat higher overall acceptances in Figure 
\ref{fig:HWacc} compared to Figure \ref{fig:acc}. This can be 
understood as due to the fact that with direct decays there are more jets 
with individually high $p_T$ than in the case of cascade 
decay chains through winos. Nevertheless, in the most compressed limit,
there is again very low sensitivity from the signals A, C, and D.

\begin{figure}[!tbp]
\mbox{\includegraphics[width=8.3cm,angle=0]{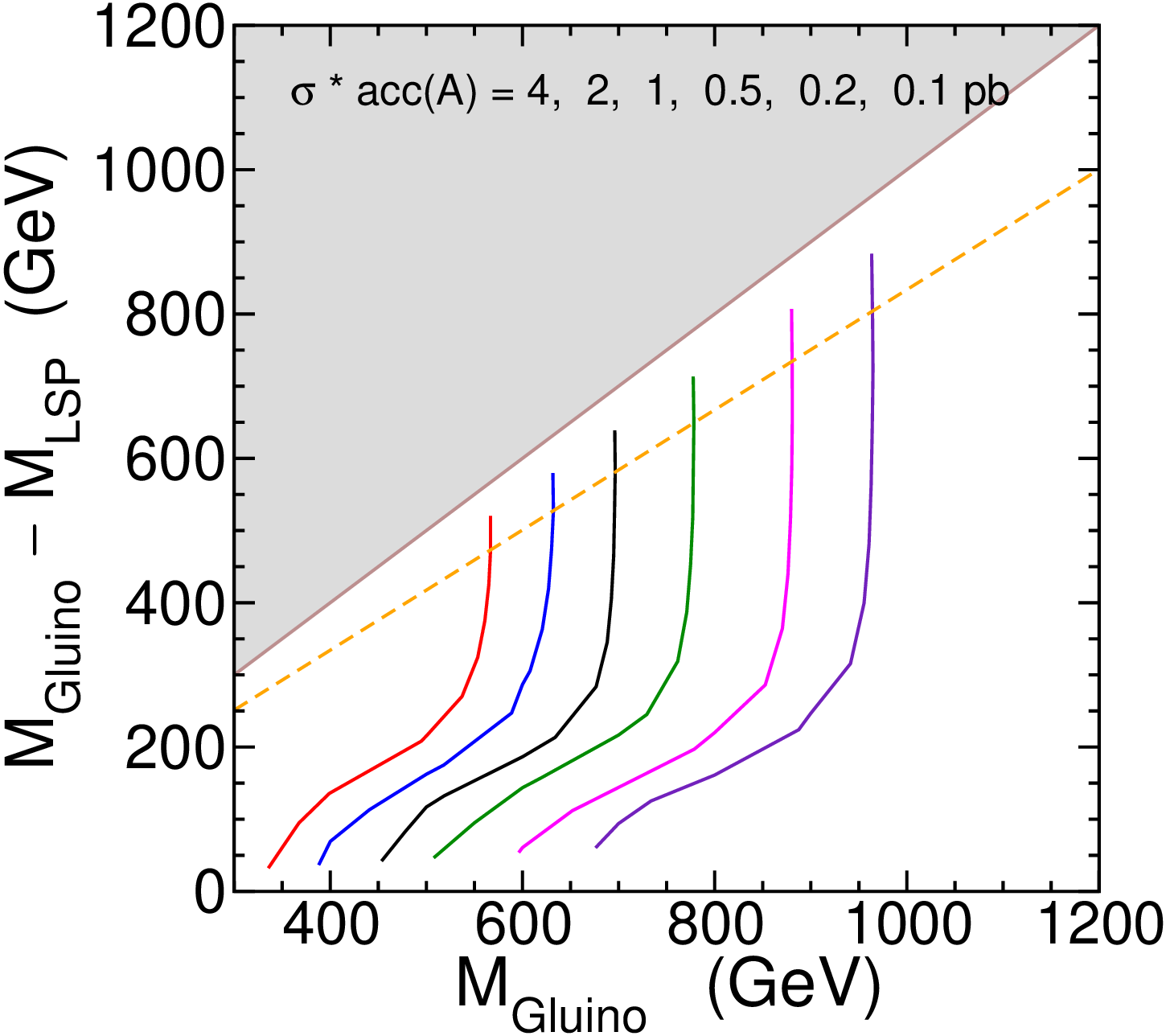}
\includegraphics[width=8.3cm,angle=0]{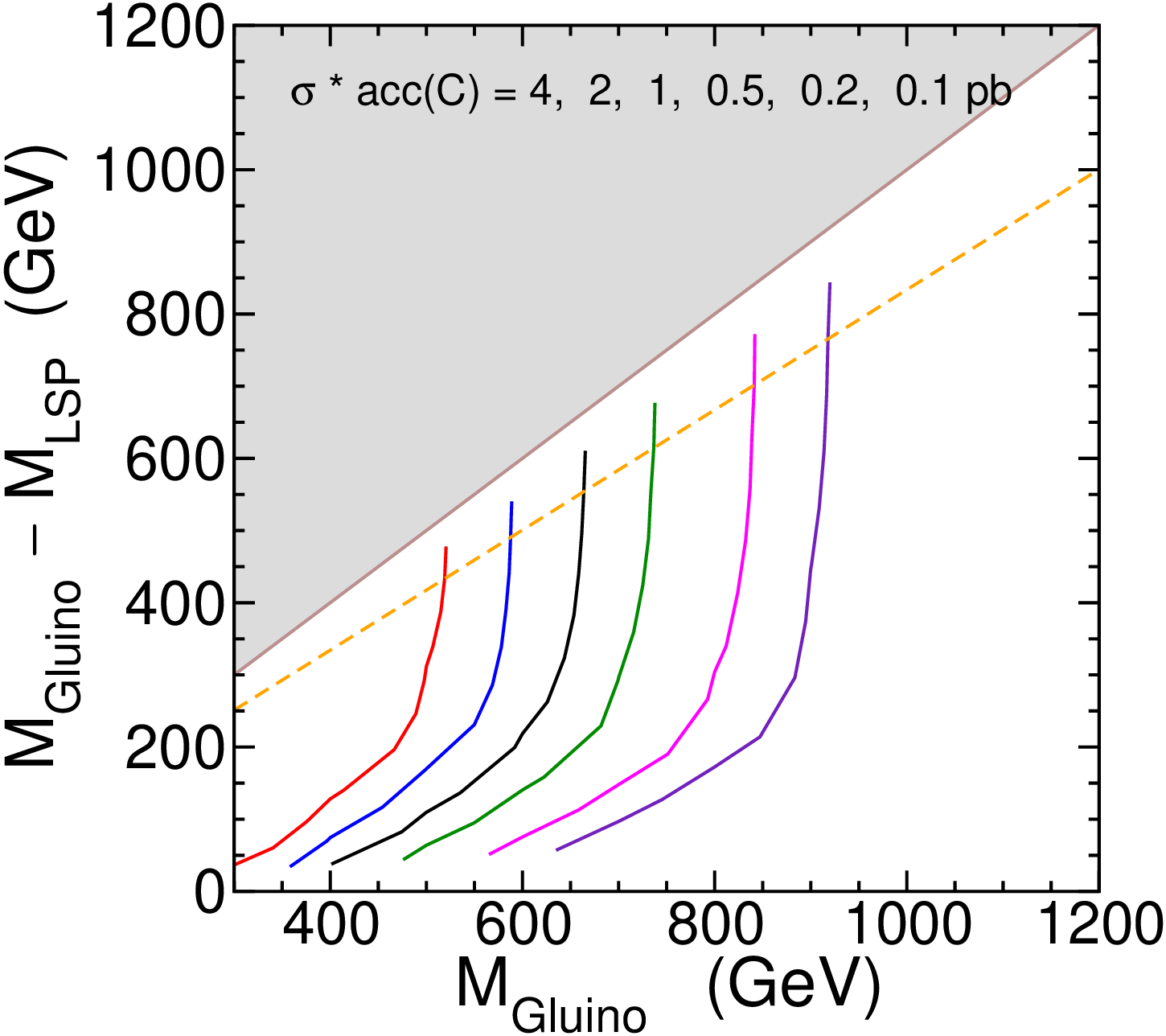}}

\vspace{0.2cm}

\mbox{
\begin{minipage}[]{8.3cm}
\includegraphics[width=8.3cm,angle=0]{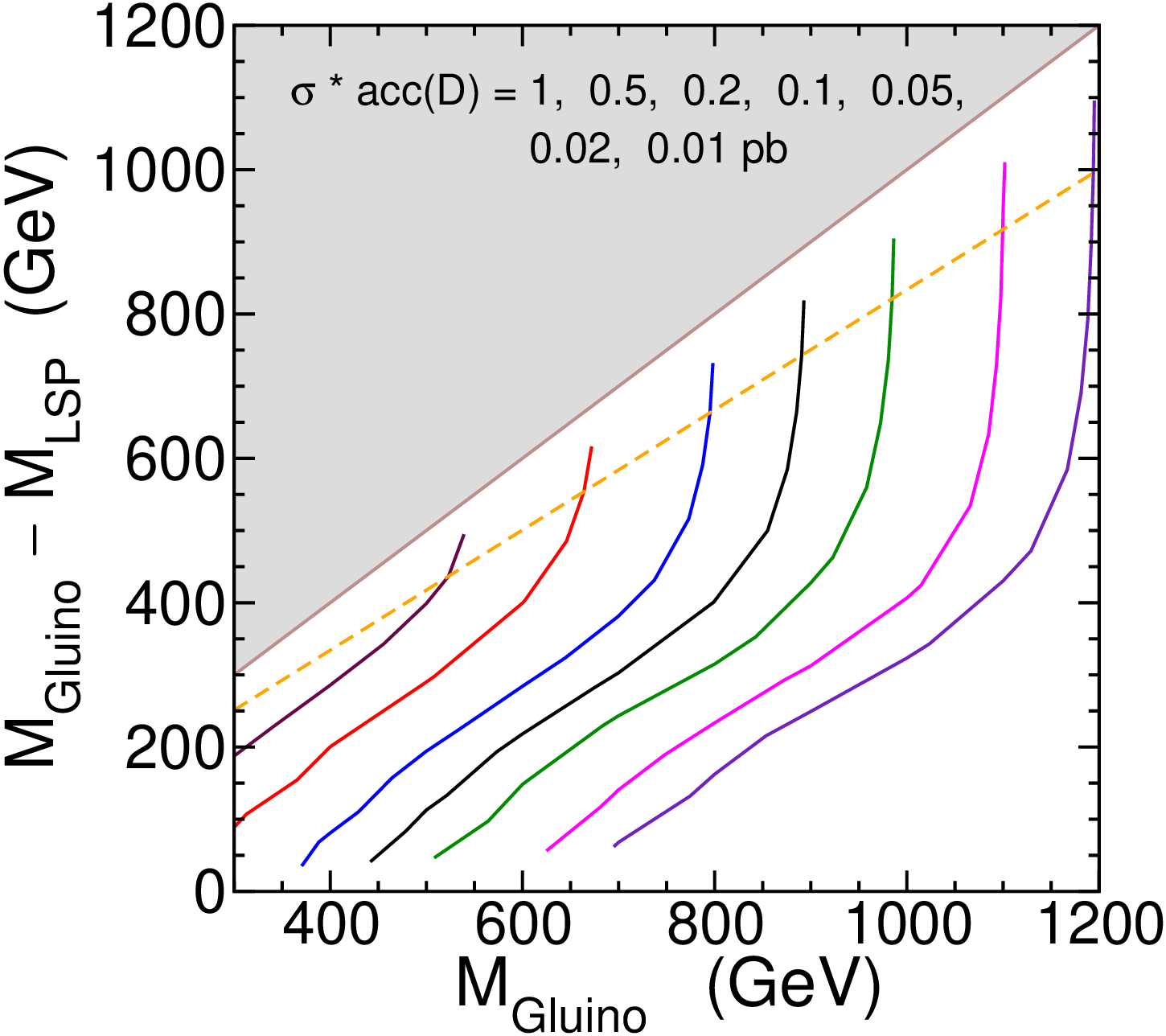}
\end{minipage}\hspace{0.6cm}
\begin{minipage}[]{7.4cm}
\caption{\label{fig:HWsigacc}
Contours of cross-section times acceptance for the heavy-wino 
models defined in section \ref{subsec:HWmodel}, in the $M_{\tilde g} - 
M_{{\tilde N_1}}$ 
vs. $M_{\tilde g}$ plane, obtained by varying the 
compression parameter $c$ between $-0.1$ and $0.9$. 
The dashed line corresponds to the mSUGRA-like case $c=0$, with 
increased compression lower in the plane.
The three panels correspond to
cuts A, C, and D.}
\phantom{xxx}
\end{minipage}
}
\end{figure}

The corresponding cross-section times acceptances are shown in Figure 
\ref{fig:HWsigacc}, in the plane of $M_{\tilde g}$ and $M_{\tilde g} - 
M_{\tilde N_1}$ as before.
The reach is slightly greater for these heavy wino models in the all-jets 
signals A,C,D than for the models in subsection \ref{subsec:cmodel}.
We note that although the greatest reach comes from cuts D when the 
compression is low, the reduction in the signal 
at high compression is less for cuts A and C. 
This is because signal D has a much stronger cut on 
$\meff$ (1000 GeV rather than 500 GeV), and again suggests that an 
intermediate value for the $\meff$ cut would yield a better reach for
compressed SUSY models. 

\clearpage

\subsection{Models with heavy squarks}
\label{subsec:HSQmodel}
\setcounter{footnote}{1}

In the models considered above, the squarks were taken to be lighter than 
the gluino. However, much heavier squarks may well be motivated by 
several factors. First, there is the LEP2 constraint on the lightest 
Higgs scalar boson mass, which increases logarithmically with the 
top-squark masses. Second, indirect constraints from  
flavor-violating and CP-violating meson decay and oscillation 
observables become weaker when squarks are heavier. Third, the so-called 
focus-point \cite{hyperbolic,focuspoint} explanation \cite{focuspointDM} 
for the WMAP-favored relic abundance of dark matter relies on having 
heavier squarks.

Therefore, in this section 
we consider a variation on the models in section \ref{subsec:cmodel},
but with all squarks taken heavy enough to essentially decouple from the 
LHC, at least for the purposes of the initial discovery process:
\beq
m_{\tilde q} &=& M_{\tilde g} + 1000 \>\, {\rm GeV}.
\eeq
The gaugino mass parameters are still related to the compression 
parameter $c$ as in eq.~(\ref{eq:cmodelgauginomasses}), so that the 
most important superpartner masses are just as depicted in Figure 
\ref{fig:cmasses} but with the squarks (including $\tilde t_1$) removed.
In these models, the most important production cross-section is from 
gluino pair production, with subsequent gluino decays $\tilde g 
\rightarrow \tilde C_1 q \bar q'$ and $\tilde N_2 q \bar q$ and $\tilde 
N_1 q \bar q$, with the first two typically dominating. The wino-like 
states then decay through on-shell or off-shell weak bosons, 
depending on the mass difference from the compression: 
$\tilde C_1 \rightarrow W^{(*)} \tilde N_1$ and $\tilde N_2 \rightarrow 
Z^{(*)} 
\tilde N_1$ or $h\tilde N_1$, with the last dominating if 
kinematically allowed.

\begin{figure}[!tbp]
\mbox{\includegraphics[width=8.3cm,angle=0]{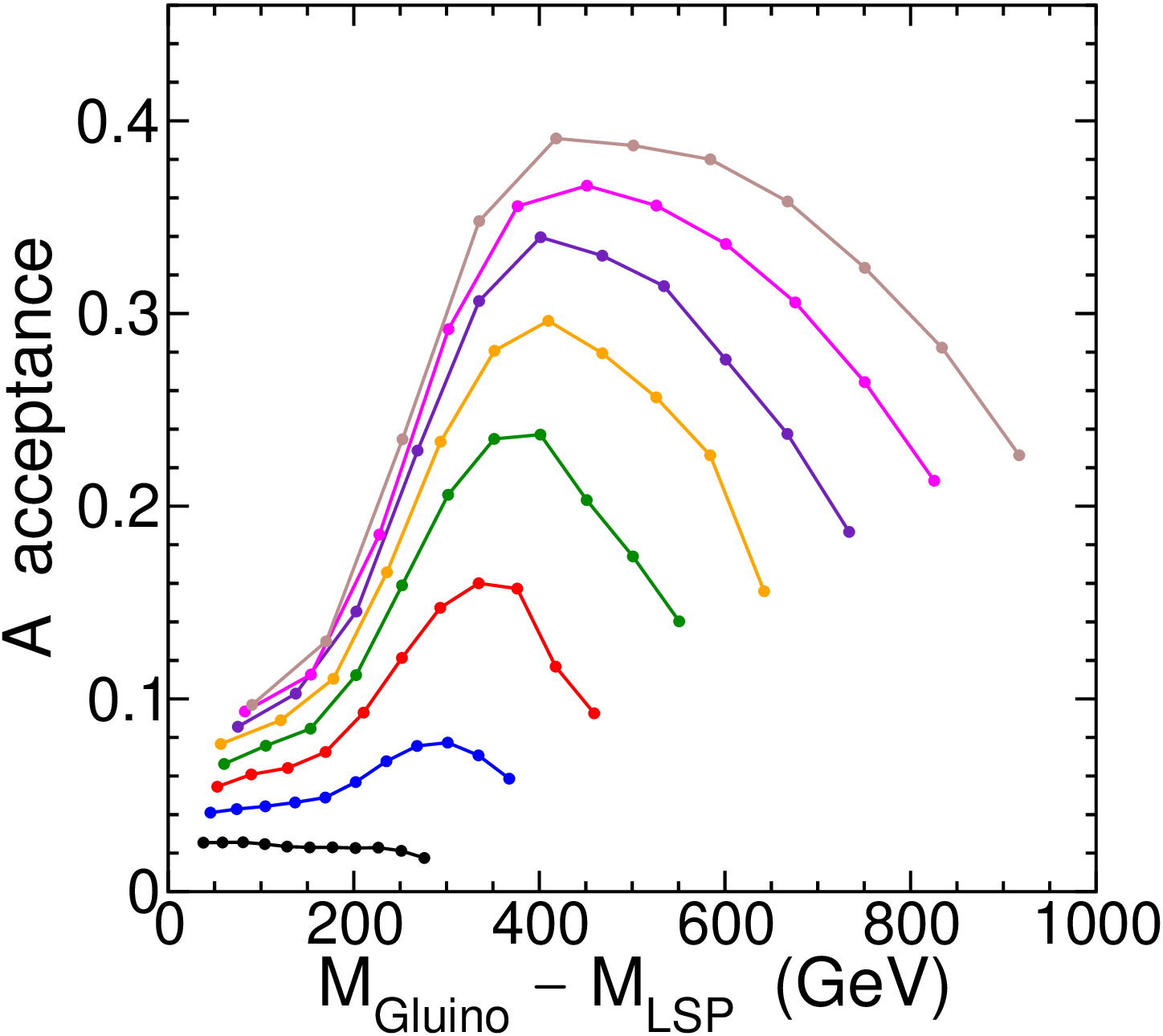}
\includegraphics[width=8.3cm,angle=0]{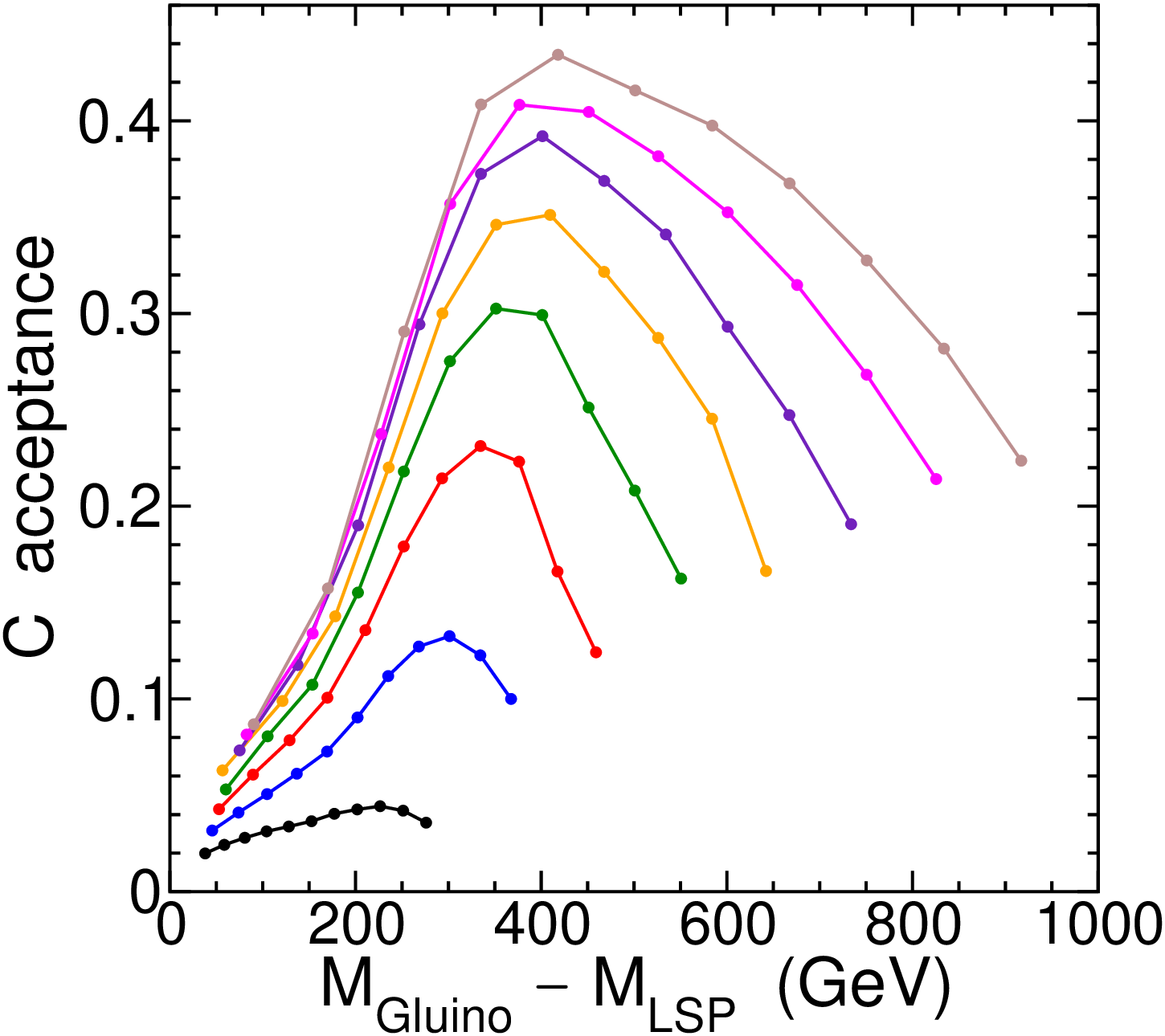}}

\vspace{0.1in}

\mbox{\includegraphics[width=8.3cm,angle=0]{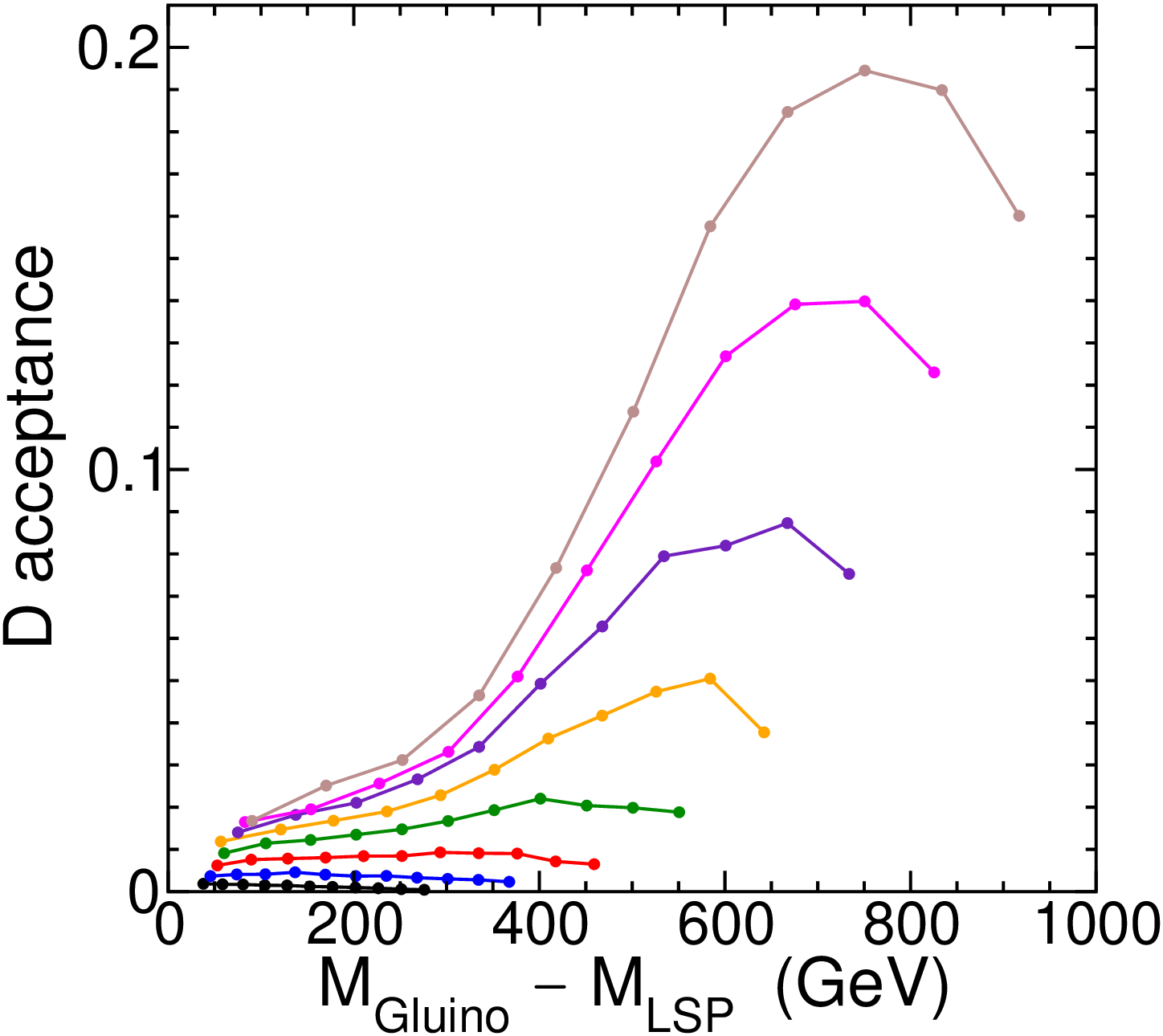}
\includegraphics[width=8.3cm,angle=0]{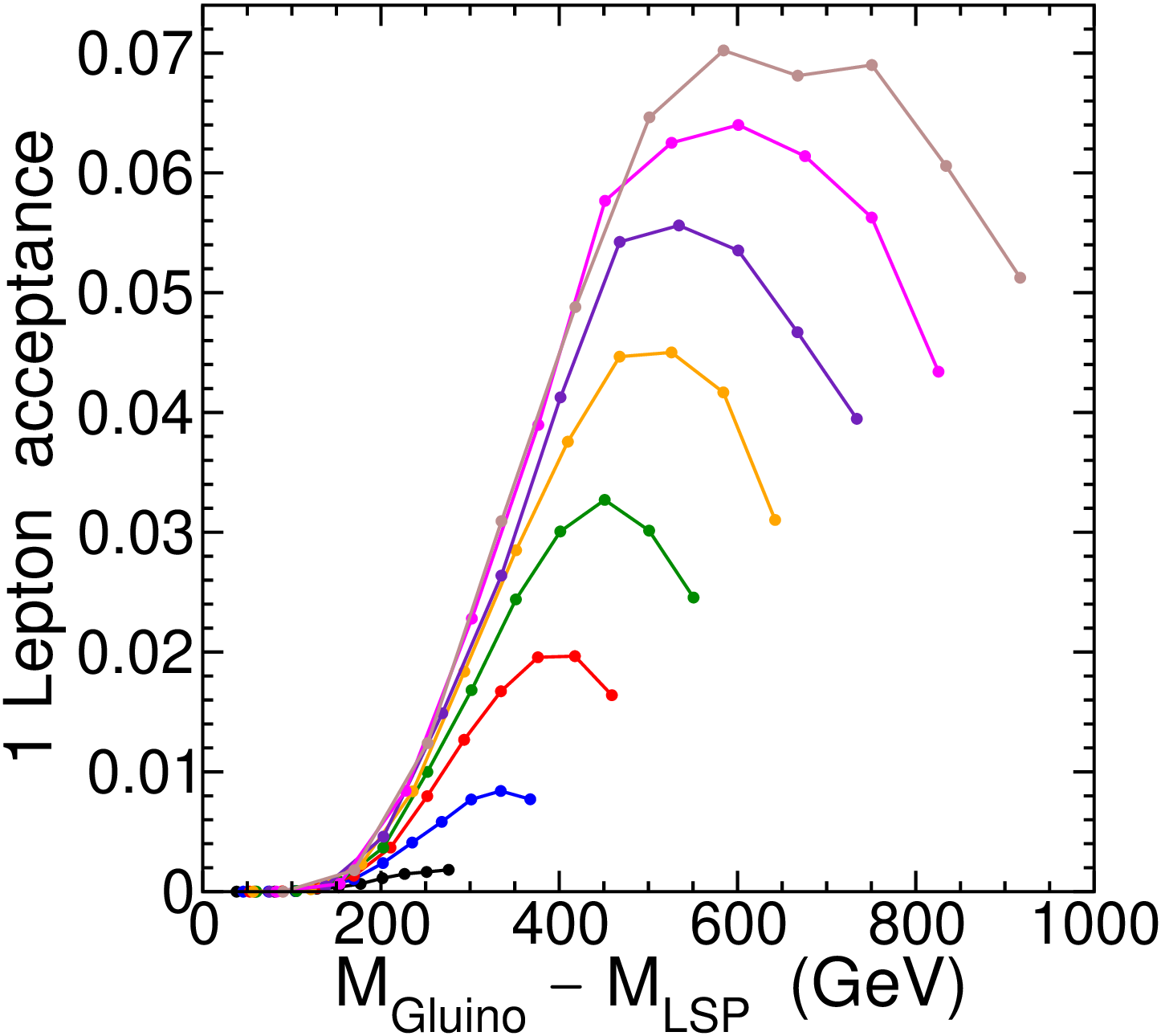}}
\caption{\label{fig:accHSQ}
The acceptances for the heavy squark 
model lines defined in section \ref{subsec:HSQmodel} 
with $M_{\tilde g} = 300, 400, \ldots, 1000$ GeV, 
as a function of $M_{\tilde g} - M_{{\tilde N_1}}$, obtained by 
varying the gaugino mass compression factor $c$. 
The dots on each line correspond to, from right to left,
$c = -0.1, 0, 0.1, \ldots 0.9$, with $c=0$ corresponding 
to the mSUGRA-like case
and $c=1$ to a completely compressed gaugino spectrum. The four panels are 
for the four sets of cuts 
A, C, D, and L.}
\end{figure}

The acceptances
for signals A, C, D, L are shown in Figure \ref{fig:accHSQ},
for $M_{\tilde g} = 300, 400, 
\ldots, 1000$ GeV, with the compression factor varying in the range
$-0.1 \leq c \leq 0.9$, as a function of the mass difference
$M_{\tilde g} - M_{{\tilde N_1}}$.
For each value of the gluino mass, the maximum acceptance occurs at
an intermediate compression factor $c$. This occurs for essentially the 
same reason
as noted in section \ref{subsec:cmodel}, namely the decays 
through winos actually have increasingly harder $\missET$ distributions
with increasing $c$, as long as $c$ is not too large, 
and increasingly softer $\meff$ distributions, leading to
more events passing the $\missET/\meff$ ratio cut. In this case, however, 
it is the gluino cascade decays that provide the effect. Since the
production is almost entirely due to gluino pairs, 
and most gluinos decay through 
winos, the rise of acceptance 
with compression for small $c$ is considerably more pronounced than in
the models of section \ref{subsec:cmodel}.

\begin{figure}[!tbp]
\mbox{\includegraphics[width=8.3cm,angle=0]{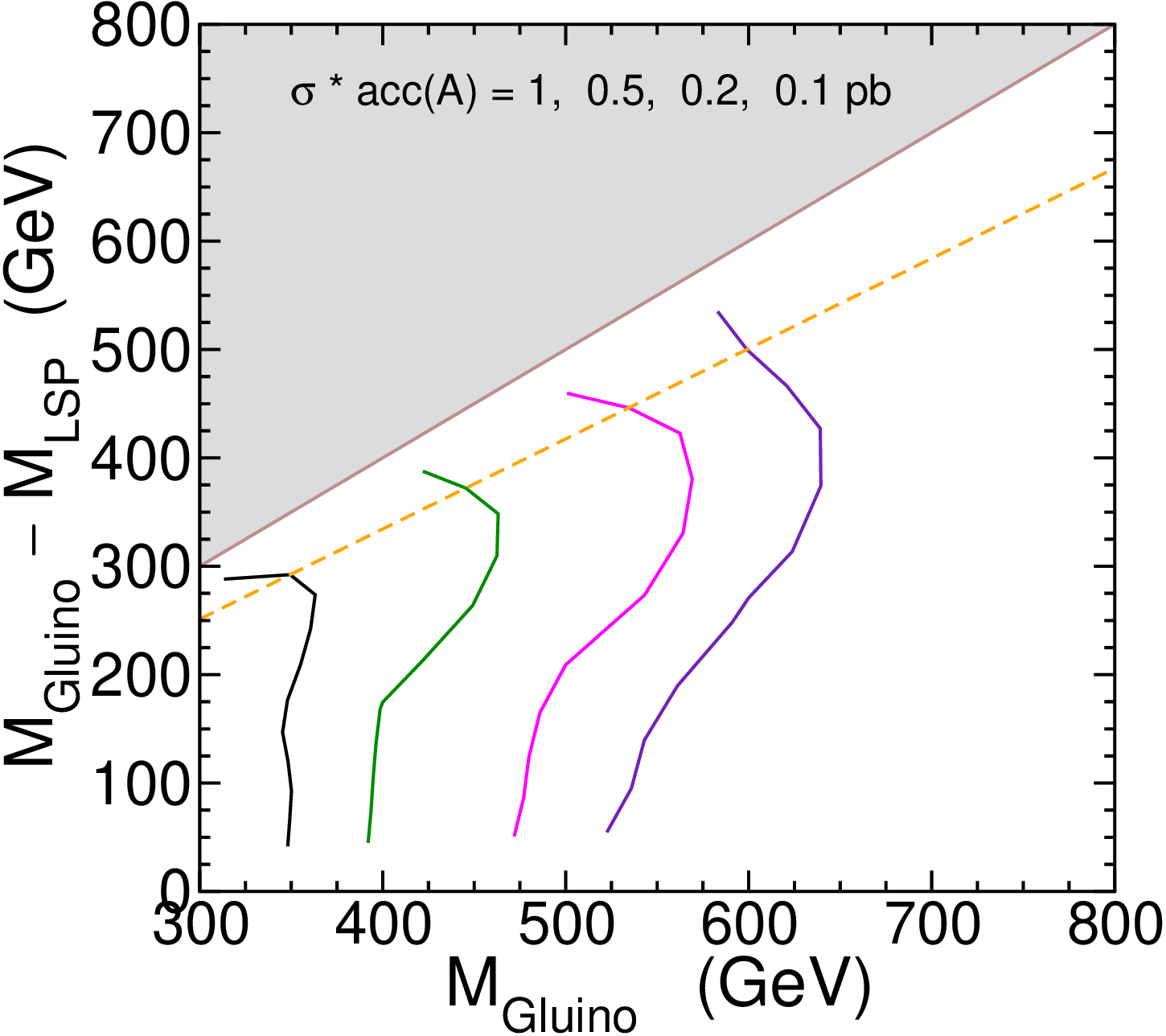}
\includegraphics[width=8.3cm,angle=0]{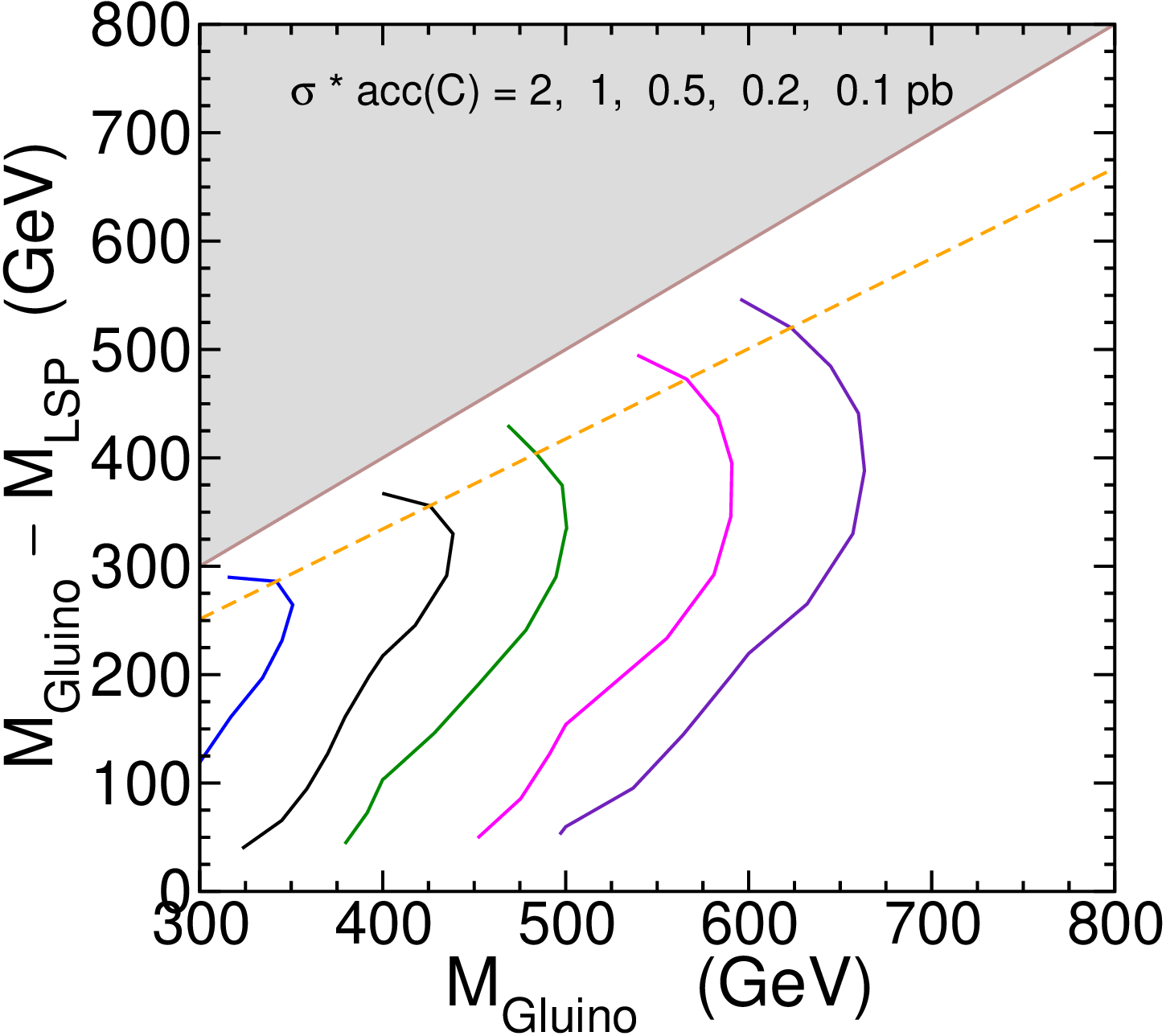}}

\vspace{0.2in}

\mbox{\includegraphics[width=8.3cm,angle=0]{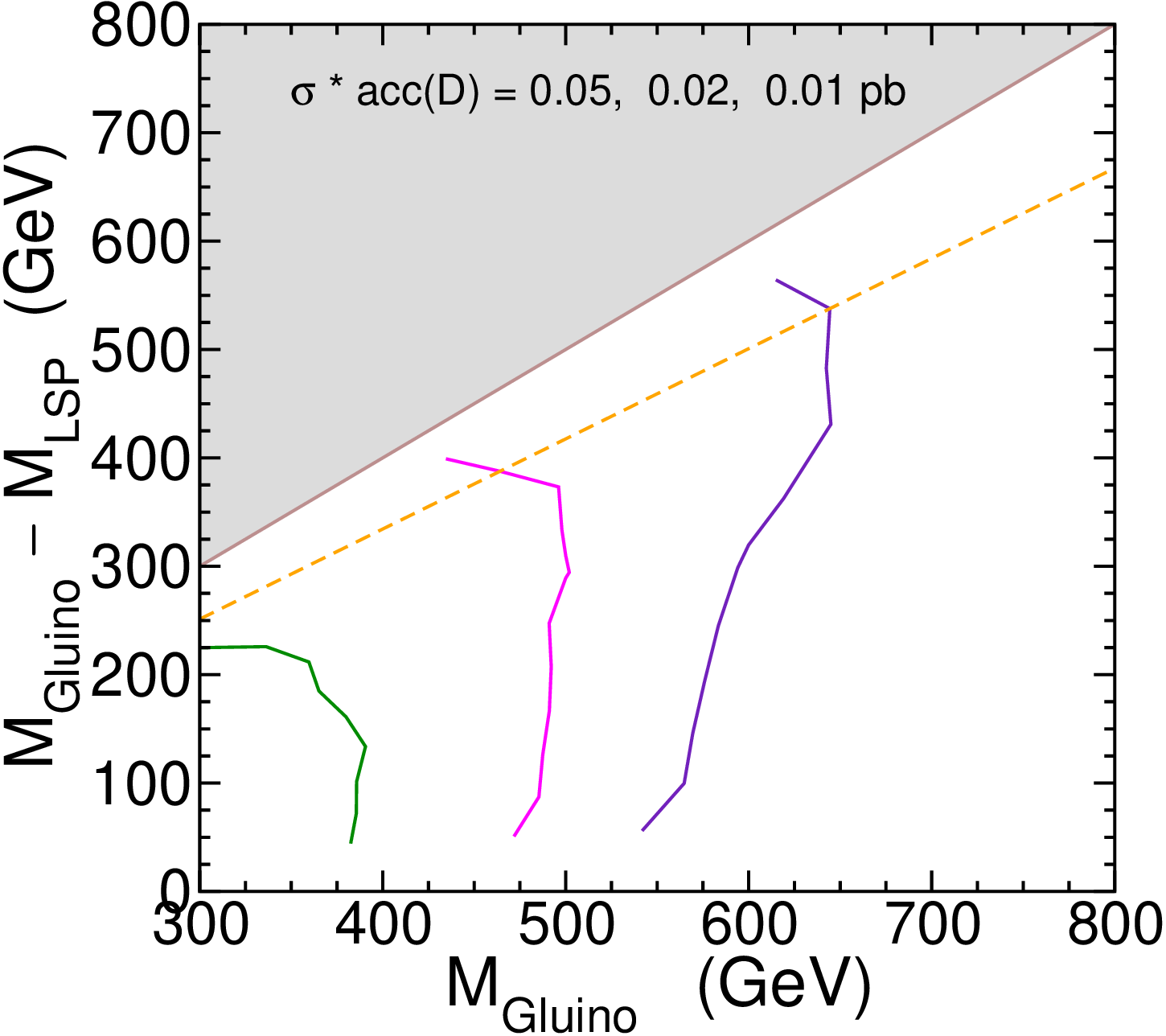}
\includegraphics[width=8.3cm,angle=0]{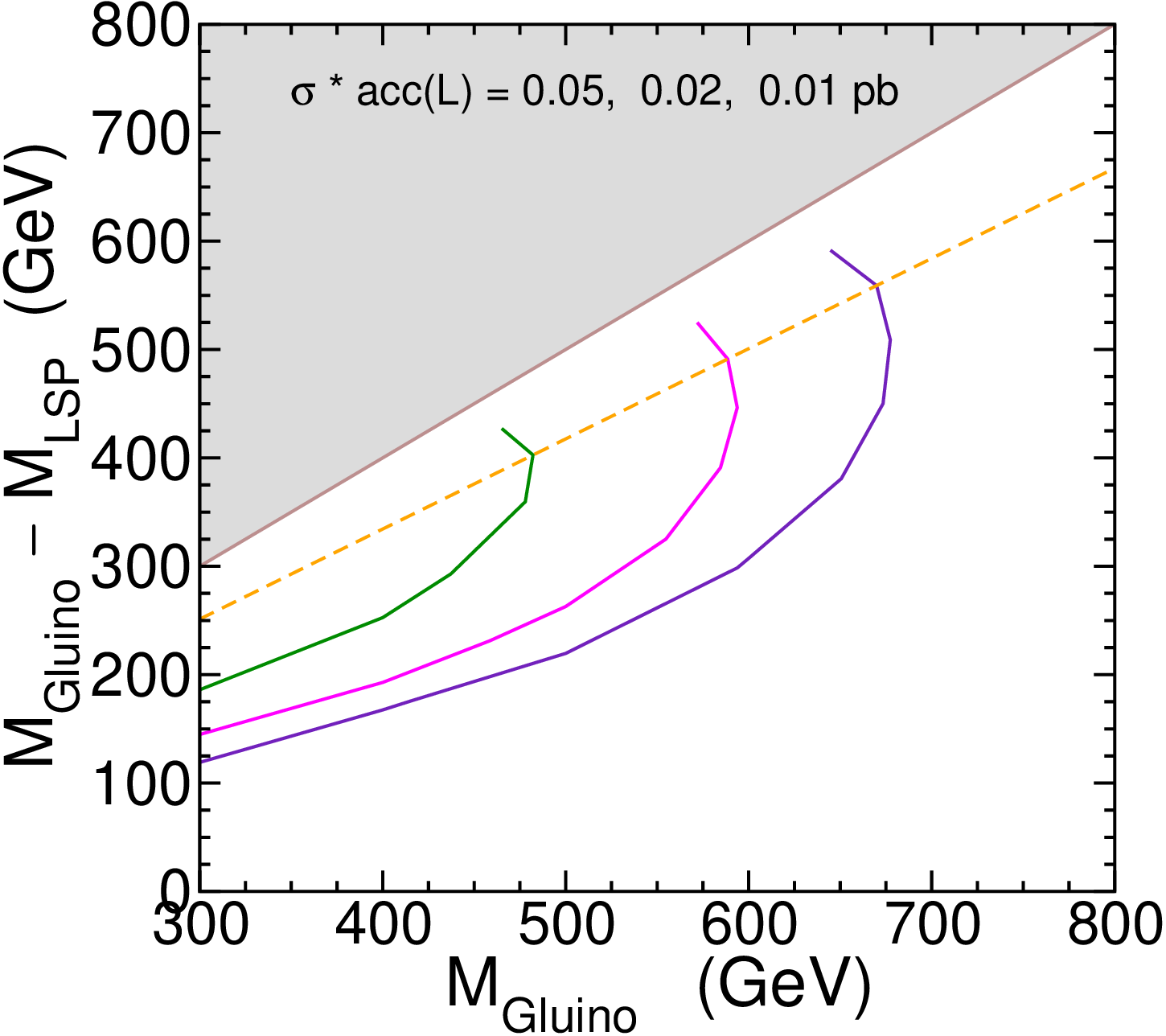}}
\caption{\label{fig:HSQsigacc}
Contours of cross-section times acceptance for the heavy-squark 
models defined in section \ref{subsec:HSQmodel}, in the $M_{\tilde g} - 
M_{{\tilde N_1}}$ 
vs. $M_{\tilde g}$ plane, obtained by varying the compression parameter 
$c$ between $-0.1$ and $0.9$. 
The dashed line corresponds to the mSUGRA-like case $c=0$, with 
increased compression lower in the plane.
The four panels correspond to
cuts A, C, D, and L.}
\end{figure}

The contours of constant cross-section times acceptance for these 
heavy squark models are shown in Figure \ref{fig:HSQsigacc}.
Because squark pair production does not make a significant contribution 
to the SUSY 
production, the reach is much weaker than in the models of the previous 
subsections. The high $\meff$ cut signal D and the single lepton signal L 
are weak here, and the latter 
turns off very quickly for higher compression factors (lower in the 
plane). In fact, only signals A and C provide any reach in the planes 
shown with $M_{\tilde g} > 300$ GeV for the 2010 data set of 
35 pb$^{-1}$ at $\sqrt{s} = 7$ TeV. Comparing with
the ATLAS limits (see Table \ref{tab:cuts}), one sees that the 2010 reach
is only up to about $M_{\tilde g} = 425$ GeV in the most favorable case,
and is less than 350 GeV for the most highly compressed case.
We also note again that a cut on $\meff$ that is intermediate 
between 
the extremes of 500 GeV (C) and 1000 GeV (D) may be more efficient in 
setting future limits or making a discovery.

\clearpage

\baselineskip=15pt

\subsection{Models with light stops motivated by dark matter}
\label{subsec:TTmodels}
\setcounter{footnote}{1}

\begin{figure}[!tbp]
\begin{flushleft}
\mbox{\includegraphics[width=8.3cm,angle=0]{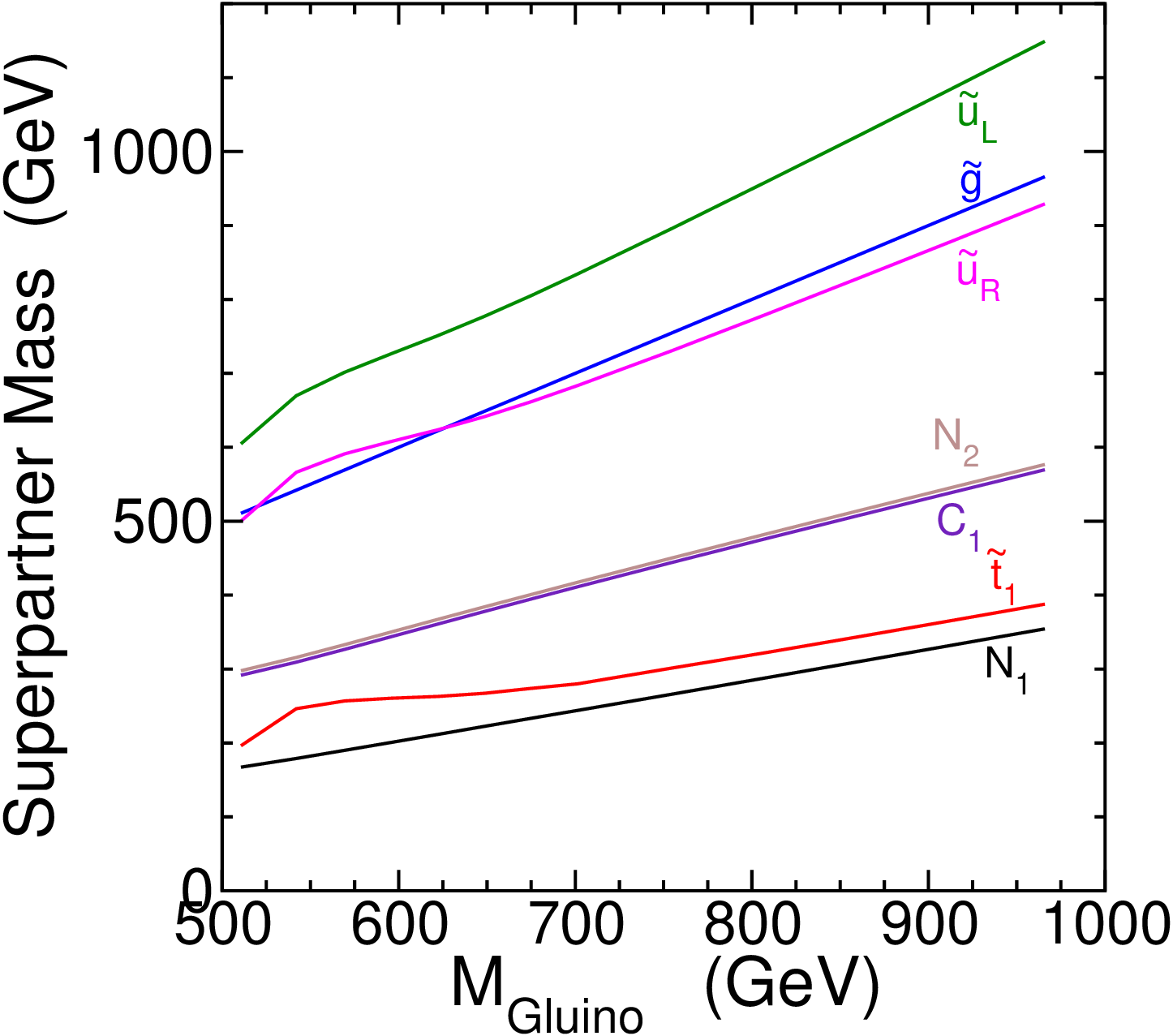}
\includegraphics[width=8.3cm,angle=0]{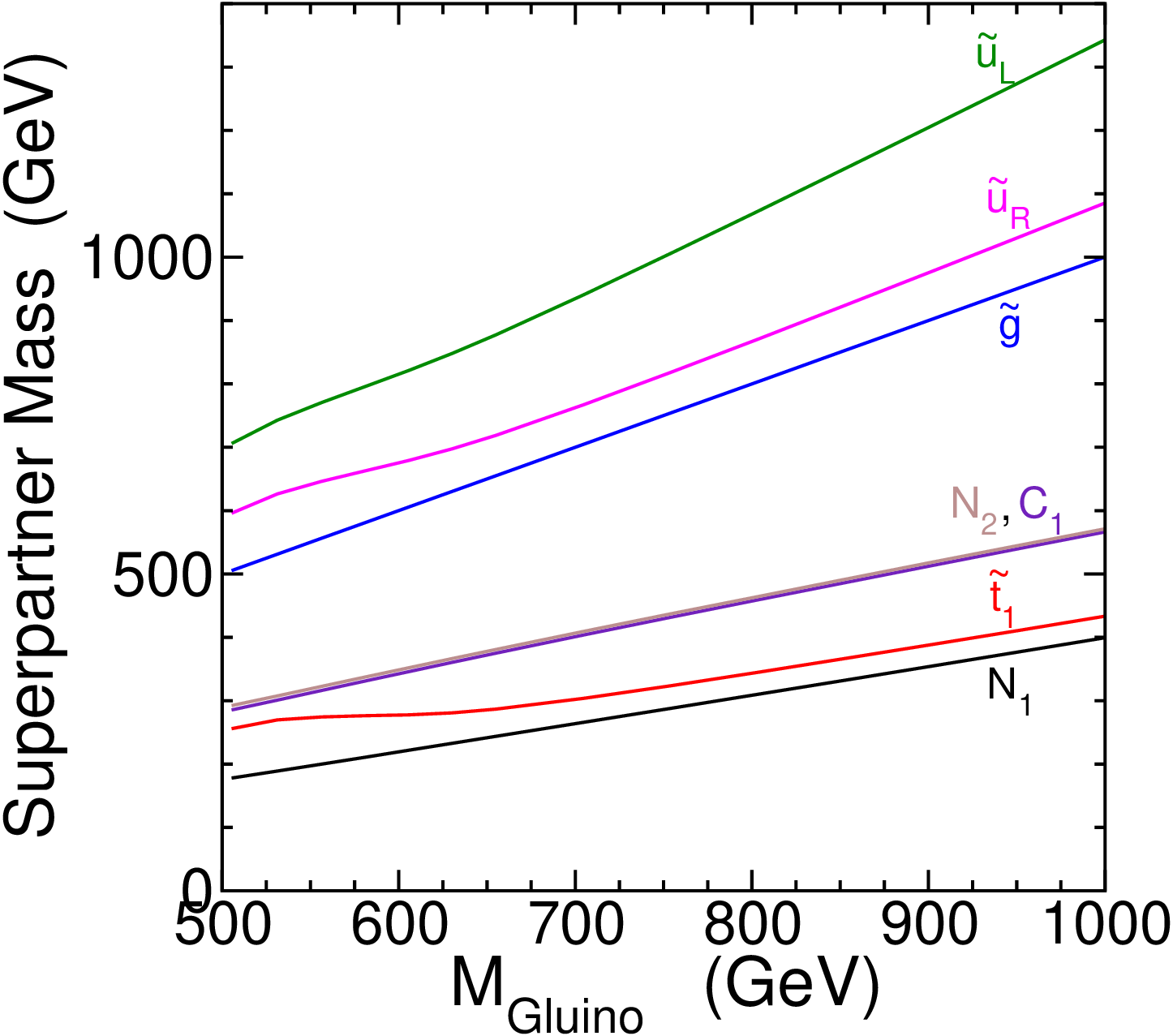}}
\caption{\label{fig:TTmasses}
The masses of important superpartners for models with $C_{24} = 0.21$
(left panel) and $C_{24} = 0.23$ (right panel), for varying $M_1 = -A_0$,
with $\tan\beta = 10$, $\mu>0$, with $m_0$ determined by requiring 
$\Omega_{\rm DM} h^2 = 0.11$.}
\end{flushleft}
\end{figure}

In this subsection, we will consider 
a class of models proposed previously in 
\cite{compressedSUSYa,compressedSUSYc} as a 
simultaneous 
solution to the supersymmetric little hierarchy problem and the problem 
of obtaining a relic density of dark matter in agreement with WMAP and 
other astrophysics data \cite{WMAP}. These models generalize mSUGRA to 
include non-universal gaugino masses in a pattern 
corresponding to an $F$-term VEV in an adjoint representation (rather 
than a singlet) of the global $SU(5)$ group that contains the Standard 
Model gauge group. Gaugino masses are parameterized by:
\beq
M_1 = m_{1/2} (1 + C_{24})\qquad
M_2 = m_{1/2} (1 + 3 C_{24})\qquad
M_3 = m_{1/2} (1 - 2 C_{24}) ,
\eeq
at $M_{\rm GUT}$, with $C_{24}=0$ corresponding to mSUGRA and
$C_{24} > 0$ to compressed SUSY. For $C_{24} \sim 0.18$ to $0.28$, the 
lighter top squark can be the next-to-lightest superpartner, and the
WMAP-favored relic abundance of dark matter is achieved by
efficient annihilations
$
\tilde N_1 \tilde N_1 \rightarrow t \overline t,
$
mediated by $\tilde t_1$ in the $t$-channel, provided that $m_t 
\lsim m_{\tilde N_1} \lsim m_t + 100$ GeV. 
(The region of parameter 
space where this occurs is 
continuously connected to the more fine-tuned case of stop coannihilation 
with dark matter.) The other parameters are as in mSUGRA; we will use
$A_0 = -M_1$, $\tan\beta = 10$, $\mu>0$, and $m_0$ chosen so as to 
obtain $\Omega_{\rm DM} h^2 = 0.11$. (In this section, we adopt a different
attitude towards the thermal dark matter density than in previous sections by 
enforcing this requirement.)
The masses of the most relevant 
superpartners are shown in Figure \ref{fig:TTmasses}, for the choices
$C_{24} = 0.21$ and 0.23.
For brevity, we chose here to consider only these two values of 
$C_{24}$, because at a
qualitative level many of the issues that impact the signal are similar
to those found in the previous sections. 

In these models, the gluino decays according to $\tilde g
\rightarrow t \tilde t_1^*$ and $\overline t \tilde t_1$ each 50\%.
In turn, the lighter top squark decays 100\% according to $\tilde t_1 
\rightarrow c \tilde N_1$, a flavor-violating 2-body decay.\footnote{It 
is also possible that the 4-body decay $\tilde t_1 \rightarrow b f 
\overline f' \tilde 
N_1$ is competitive \cite{stopdecays}. We assume that there is sufficient 
flavor violation 
in the soft supersymmetry breaking sector to assure that the 2-body decay 
wins and is prompt.} Although the pair production of light stops has the 
largest 
cross-section of all SUSY processes at the LHC, it has a very low acceptance for all 
signals considered here, due to the small kinematic phase space of this 
decay; the charm jets typically have very low $p_T$. 
The first- and second-family squarks decay mostly according to $\tilde 
q_L \rightarrow q \tilde g$ and $q_R \rightarrow q \tilde N_1$, with 
subdominant decays to the neutralinos and charginos $\tilde N_2, \tilde 
N_3$ and $\tilde C_1$, which are higgsino-like in these models.
(The wino-like states are heavier and essentially decouple.)

The compression of the spectrum in these dark-matter motivated models is 
not extreme, with a ratio
$M_{\tilde g}/M_{\tilde N_1}$ of roughly 2.7 for the $C_{24} = 0.21$ 
model line and 2.9 for the $C_{24} = 0.23$ model line, corresponding 
roughly to compression factors of $c=0.21$ and $c=0.25$ in the language 
of section \ref{subsec:cmodel}. The most striking qualitatively 
different feature 
is the presence of
top quarks in all decays involving the gluino.\footnote{Since the gluino 
is 
Majorana, the charges of the two top quarks in each event are 
uncorrelated, leading to a rare but very low-background 
same-sign top-quark signal, not explored here.
See for example refs.~\cite{Kraml:2005kb} and \cite{compressedSUSYc}.} 

\begin{figure}[!tbp] \begin{flushleft}
\mbox{\includegraphics[width=8.3cm,angle=0]{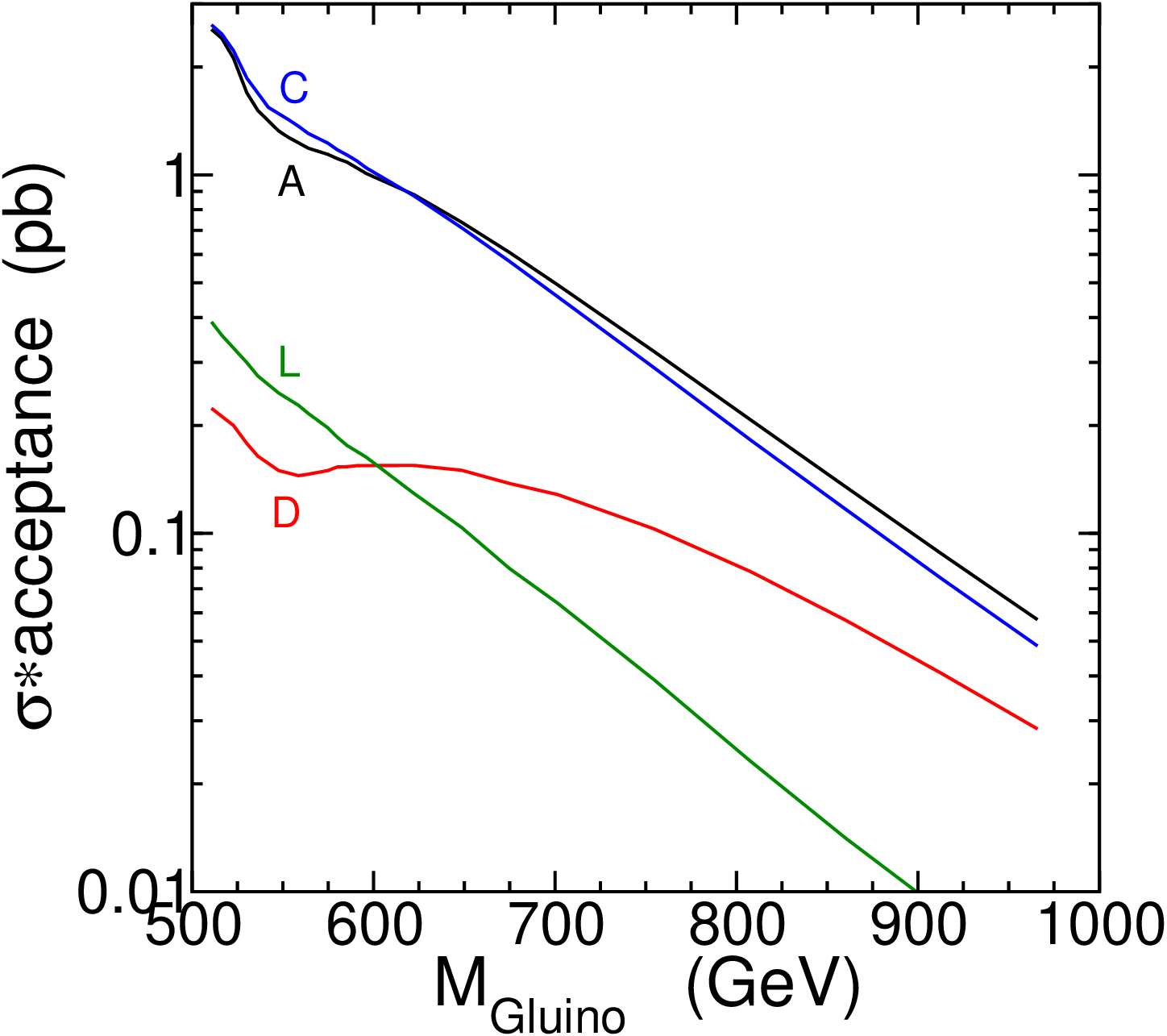}
\includegraphics[width=8.3cm,angle=0]{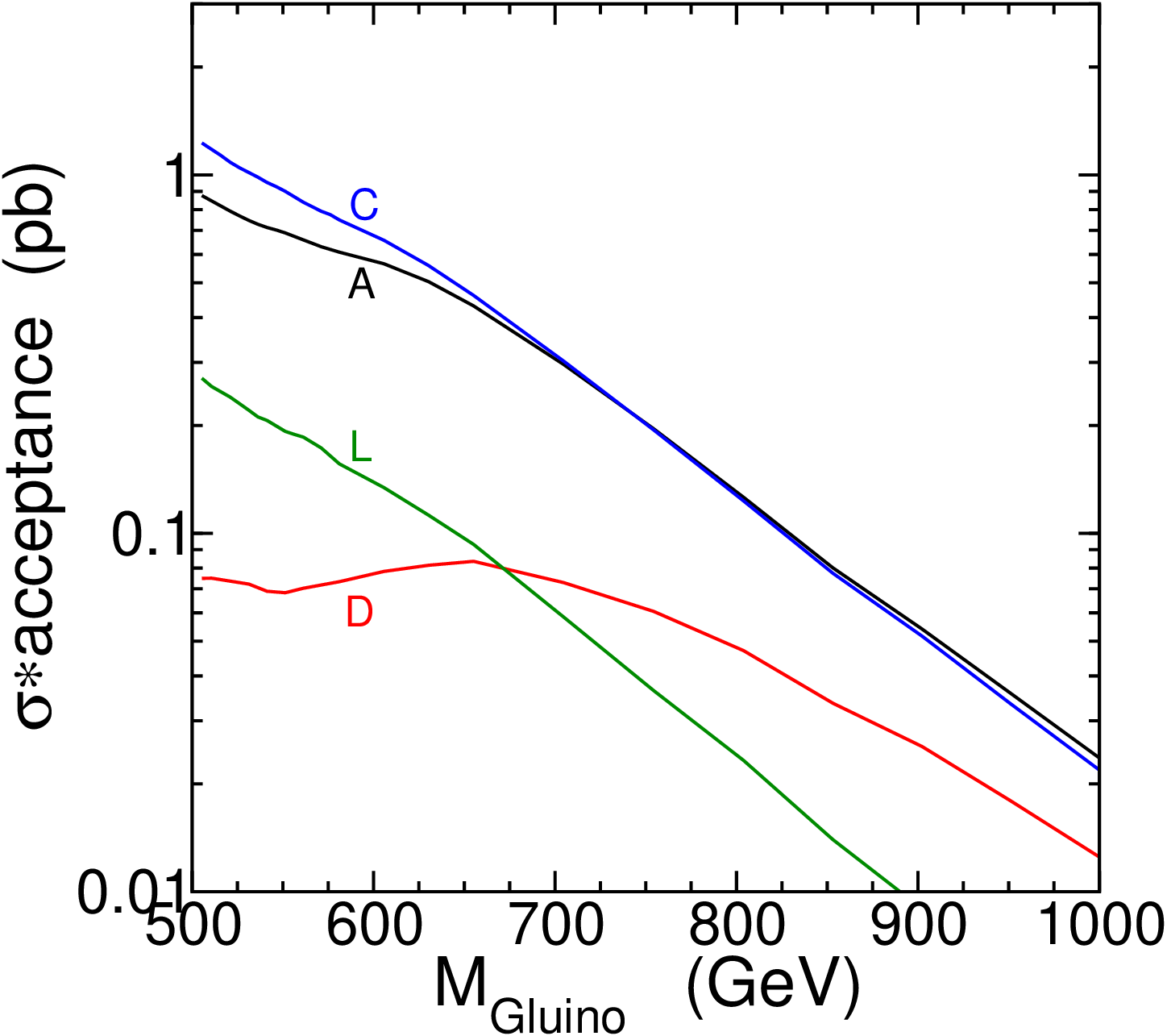}}
\caption{\label{fig:TTsigacc}
The cross-section times acceptance for the same model lines in Figure
\ref{fig:TTmasses}, as a function of the gluino mass, 
for the four sets of cuts 
A, C, D, and L.}
\end{flushleft}
\end{figure}

The cross-section times acceptance for signals A, C, D, and L for these 
model lines is shown in 
Figure \ref{fig:TTsigacc} as a function of the gluino mass.
As might be expected from the results of the previous section, the 
moderate compression of the spectrum in these examples is not enough to 
strongly deplete the signal. For the $C_{24} = 0.21$ model line, the 2010 
ATLAS bounds in Table \ref{tab:cuts} should exclude up to
$M_{\tilde g} = 575$ GeV with signal C (although this should be considered
approximate in the absence of a dedicated study of backgrounds and detectors
responses), and perhaps about the same from 
signal L (taking into account the fact, mentioned at the end of section 
\ref{sec:procedures} above, that our methodology tends to give 20\% to 
50\% larger acceptances for the 1-lepton channel than were reported by 
ATLAS). The exclusion obtainable from signal D is hard to estimate, as 
the cross-section times acceptance is relatively flat over a large range 
of gluino masses, but it may already be above $M_{\tilde g} = 725$ GeV 
with 2010 data. Obtaining a real estimate of the exclusion would require 
a dedicated analysis of the backgrounds and detector responses. The 
$C_{24} = 0.23$ model line has weaker signals, due to having heavier 
squarks. Here, the signal D is not quite able to exclude any models along 
the line, and the signal L reach is the best but with 2010 data does not 
extend much beyond $M_{\tilde g} = 600$ GeV.

Fortunately, with the greater integrated luminosity to become available 
soon, these models will be confronted out to much larger gluino masses. 
For high masses, the signal D appears to have the best sensitivity, out 
of the ones considered here, but it is likely that modified sets of cuts will 
do even better. In particular, the large multiplicity of jets in the two 
top decays suggests that $\meff$ would be better replaced by a variable 
that summed over more jets, to better capture the distinction between 
these events and QCD backgrounds. Also, bottom tagging could be helpful.

\section{Outlook}
\label{sec:conclusion}
\setcounter{equation}{0}
\setcounter{footnote}{1}

In this paper, we have studied the reach of the $\sqrt{s} = 7$ TeV LHC 
for supersymmetric models with compressed mass spectra. Such models are 
not just interesting ways to hide SUSY, but can be motivated as providing 
a solution to the supersymmetric little hierarchy problem. We found that 
with mild to moderate compression, the acceptances are good and sometimes 
even much better than in mSUGRA. Acceptances drastically decrease for more 
severe compression, as expected due to the low visible energy in 
each event. The $\meff$ cut is typically more damaging to the signal than 
the $\missET$ cut, but even for the most extreme compression there is 
some reach in the multi-jet channels. For high compression, the 1-lepton 
signal goes away 
completely, due to the difficulty in getting high $p_T$ leptons and 
large transverse mass $m_T$ from decays with small mass differences.

These studies suggest that signals with an intermediate (between the 
extremes of 500 and 1000 GeV) cut on $\meff$ might be more useful in 
extending the reach. Also, signals that cut on kinematic variables that 
augment $\meff$ by involving more than 3 jets may be better at probing 
signals from models with 3-body and cascade decays. A question that 
should be addressed in future analyses is whether compressed SUSY might 
contribute to QCD background control regions (used to estimate background 
from data) in dangerous way. We look forward to the data currently being 
gathered at the LHC probing supersymmetry in large regions of parameter 
space by using comprehensive search strategies involving a variety of 
different signals.

{\em Acknowledgments:} We thank Johan Alwall, Jerry Blazey, Jay Wacker, 
and Adam Yurkewicz for useful conversations.
The work of TJL was supported in part by the U.S. Department of Energy, 
Division of High Energy Physics, under Contract DE-AC02-06CH11357.
The work of SPM was supported in part by the 
National Science Foundation grant number PHY-0757325.



\begin{thebibliography}{90}
\baselineskip=10.3pt

\bibitem{ATLASTDR}
The ATLAS collaboration,
``ATLAS Detector and physics performance technical design report", Volume 2.
CERN-LHCC-99-15, ATLAS-TDR-15, May 1999.

\bibitem{CMSTDR}
The CMS collaboration,
``CMS Physics Technical Design Report", Volume 1, Detector Performance
and Software.
CERN-LHCC-2006-001, CMS TDR 8.1, February 2006.

\bibitem{SUSYreviews}
For reviews of supersymmetry at the TeV scale, see
  S.P.~Martin,
  ``A supersymmetry primer,''
  [hep-ph/9709356] (version 5, December 2008).
  M.~Drees, R.~Godbole and P.~Roy,
  ``Theory and phenomenology of sparticles: An account of four-dimensional N=1
  supersymmetry in high energy physics,''
{\it  World Scientific (2004).}
%
  H.~Baer and X.~Tata,
  ``Weak scale supersymmetry: From superfields to scattering events,''
{\it  Cambridge University Press (2006)}.



\bibitem{ATLASL}
  G.~Aad {\it et al.} [ Atlas Collaboration ],
  ``Search for supersymmetry using final states with one lepton, jets, 
and missing transverse momentum with the ATLAS detector in sqrt{s} = 7 
TeV pp,'' [1102.2357 [hep-ex]].

\bibitem{ATLASJ}
  G.~Aad {\it et al.} [ Atlas Collaboration ],
  ``Search for squarks and gluinos using final states with jets and 
missing transverse momentum with the ATLAS detector in sqrt(s) = 7 TeV 
proton-proton collisions,''
[1102.5290 [hep-ex]].

\bibitem{Aad:2011ks}
  G.~Aad {\it et al.} [ ATLAS Collaboration ],
  ``Search for supersymmetry in pp collisions at sqrt{s} = 7TeV in final 
states with missing transverse momentum and b-jets,''
[1103.4344 [hep-ex]].

\bibitem{Aad:2011xm}
  G.~Aad {\it et al.} [ ATLAS Collaboration ],
  ``Search for supersymmetric particles in events with lepton pairs and 
large missing transverse momentum in $\sqrt{s}=7$ TeV proton-proton 
collisions with the ATLAS experiment,''
[1103.6214 [hep-ex]].


\bibitem{Khachatryan:2011tk}
  V.~Khachatryan {\it et al.}  [CMS Collaboration],
  ``Search for Supersymmetry in pp Collisions at 7 TeV in Events with Jets and
  Missing Transverse Energy,''
  Phys.\ Lett.\  B {\bf 698}, 196 (2011)
  [1101.1628 [hep-ex]].

\bibitem{Chatrchyan:2011bz}
  S.~Chatrchyan {\it et al.}  [CMS Collaboration],
  ``Search for Physics Beyond the Standard Model in Opposite-Sign Dilepton
  Events at $\sqrt{s} = 7$ TeV,''
  1103.1348 [hep-ex].

\bibitem{Chatrchyan:2011wb}
  S.~Chatrchyan {\it et al.}  [CMS Collaboration],
  ``Search for new physics with same-sign isolated dilepton events with jets
  and missing transverse energy at the LHC,''
  1104.3168 [hep-ex].



\bibitem{Baer:1986ki}
  H.~Baer, D.~Karatas and X.~Tata,
  Phys.\ Lett.\  B {\bf 183}, 220 (1987).


\bibitem{Baer:2007uz}
  H.~Baer, A.~Box, E.K.~Park and X.~Tata,
  JHEP {\bf 0708}, 060 (2007)
  0707.0618 [hep-ph].

\bibitem{Alves:2010za}
D.S.M.~Alves, E.~Izaguirre and J.G.~Wacker,
  ``It's On: Early Interpretations of ATLAS Results in Jets and Missing Energy
  Searches,''
  1008.0407 [hep-ph];

\bibitem{Conley:2010du}
  J.A.~Conley, J.S.~Gainer, J.L.~Hewett, M.P.~Le and T.G.~Rizzo,
  ``Supersymmetry without prejudice at the LHC,''
  1009.2539 [hep-ph];
  ``Supersymmetry without prejudice at the 7 TeV LHC,''
  1103.1697 [hep-ph].

\bibitem{Scopel:2011qt}
S.~Scopel, S.~Choi, N.~Fornengo and A.~Bottino,
  1102.4033 [hep-ph].

\bibitem{Buchmueller:2011aa}
  O.~Buchmueller {\it et al.},
  Eur.\ Phys.\ J.\  C {\bf 71}, 1634 (2011)
  1102.4585 [hep-ph].
See also 
P.~Bechtle {\it et al.},
  arXiv:1102.4693 [hep-ph]
for a similar study within mSUGRA.

\bibitem{Alves:2011sq}
D.S.M.~Alves, E.~Izaguirre and J.G.~Wacker,
  ``Where the Sidewalk Ends: Jets and Missing Energy Search Strategies for the
  7 TeV LHC,''
  1102.5338 [hep-ph].

\bibitem{Akula:2011dd}
  S.~Akula, D.~Feldman, Z.~Liu, P.~Nath and G.~Peim,
  ``New Constraints on Dark Matter from CMS and ATLAS Data,''
  1103.5061 [hep-ph].



\bibitem{MGME} 
J.~Alwall {\it et al.},
  JHEP {\bf 0709}, 028 (2007) 
  F.~Maltoni and T.~Stelzer,  
  ``MadEvent: Automatic event generation with MadGraph,''
  JHEP {\bf 0302}, 027 (2003) 
  [hep-ph/0208156],  
  T.~Stelzer and W.~F.~Long,
  Comput.\ Phys.\ Commun.\  {\bf 81}, 357 (1994)
  [hep-ph/9401258].

\bibitem{CTEQ}
J.~Pumplin {\it et al.}, 
  JHEP {\bf 0207}, 012 (2002)
  [hep-ph/0201195].

\bibitem{Pythia}  
T.~Sjostrand, S.~Mrenna and P.~Skands,  
  JHEP {\bf 0605}, 026 (2006)  
  [hep-ph/0603175].  

\bibitem{PGS}
John Conway {et al.,} ``PGS4: Pretty Good Simulation of high energy collisions",
\hfill\\
\verb+http://www.physics.ucdavis.edu/~conway/research/software/pgs/pgs4-general.htm+.

\bibitem{matching}
\verb+http://cp3wks05.fynu.ucl.ac.be/twiki/bin/view/Main/IntroMatching+.
See also
J.~Alwall, S.~de Visscher and F.~Maltoni,
  JHEP {\bf 0902}, 017 (2009)
  0810.5350 [hep-ph].
J.~Alwall {\it et al.},
  Eur.\ Phys.\ J.\  C {\bf 53}, 473 (2008)
  0706.2569 [hep-ph].

\bibitem{prospino}
Prospino 2.1, available at 
\verb+http://www.ph.ed.ac.uk/~tplehn/prospino/+,
uses results found in:
W.~Beenakker, R.~Hopker, M.~Spira and P.M.~Zerwas, 
   Nucl.\ Phys.\  B {\bf 492}, 51 (1997)
   [hep-ph/9610490],
W.~Beenakker et al, 
   Nucl.\ Phys.\  B {\bf 515}, 3 (1998)  
   [hep-ph/9710451],  
W.~Beenakker et al, 
   Phys.\ Rev.\ Lett.\  {\bf 83}, 3780 (1999)
   [Erratum-ibid.\  {\bf 100}, 029901 (2008)]  
   [hep-ph/9906298],
M.~Spira,  
   [hep-ph/0211145],  
T.~Plehn,  
   Czech.\ J.\ Phys.\  {\bf 55}, B213 (2005)  
   [hep-ph/0410063].  


\bibitem{compressedSUSYa}
  S.P.~Martin,
  Phys.\ Rev.\  D {\bf 75}, 115005 (2007)
  [hep-ph/0703097],
%
  Phys.\ Rev.\  D {\bf 76}, 095005 (2007)
  [hep-ph/0707.2812].

\bibitem{compressedSUSYc}
  S.P.~Martin,
  Phys.\ Rev.\  {\bf D78}, 055019 (2008).
  0807.2820 [hep-ph].

\bibitem{WMAP}
E.~Komatsu {\it et al.}  [WMAP Collaboration],
  Astrophys.\ J.\ Suppl.\  {\bf 192}, 18 (2011)
  1001.4538 [astro-ph.CO].

\bibitem{KaneKing}
  G.L.~Kane and S.F.~King,
  Phys.\ Lett.\  B {\bf 451}, 113 (1999)
  [hep-ph/9810374],
  M.~Bastero-Gil, G.L.~Kane and S.F.~King,
  Phys.\ Lett.\  B {\bf 474}, 103 (2000)
  [hep-ph/9910506].

\bibitem{hyperbolic}
  K.L.~Chan, U.~Chattopadhyay and P.~Nath,
  Phys.\ Rev.\ D {\bf 58}, 096004 (1998)
  [hep-ph/9710473].

\bibitem{focuspoint}
J.L.~Feng, K.T.~Matchev and T.~Moroi,
  Phys.\ Rev.\ Lett.\  {\bf 84}, 2322 (2000)
  [hep-ph/9908309];
  Phys.\ Rev.\ D {\bf 61}, 075005 (2000)
  [hep-ph/9909334].

\bibitem{focuspointDM}
J.L.~Feng, K.T.~Matchev and F.~Wilczek,
  Phys.\ Lett.\ B {\bf 482}, 388 (2000)
  [hep-ph/0004043].

\bibitem{stopdecays}
K.-i.~Hikasa, M.~Kobayashi,
  Phys.\ Rev.\  {\bf D36}, 724 (1987)
C.~Boehm, A.~Djouadi and Y.~Mambrini,
  Phys.\ Rev.\  D {\bf 61}, 095006 (2000)
  [hep-ph/9907428].
S.P.~Das, A.~Datta and M.~Guchait,
  Phys.\ Rev.\  D {\bf 65}, 095006 (2002)
  [hep-ph/0112182].
G.~Hiller and Y.~Nir,
  JHEP {\bf 0803}, 046 (2008)
  [0802.0916 [hep-ph]].
A.D.~Box and X.~Tata,
  Phys.\ Rev.\  D {\bf 79}, 035004 (2009)
  [Erratum-ibid.\  D {\bf 82}, 119905 (2010)]
  0810.5765 [hep-ph].
M.~Muhlleitner and E.~Popenda,
  JHEP {\bf 1104}, 095 (2011)
  1102.5712 [hep-ph].

\bibitem{Kraml:2005kb}
  S.~Kraml and A.R.~Raklev,
  Phys.\ Rev.\  D {\bf 73}, 075002 (2006)
  [hep-ph/0512284].

\end{thebibliography}
\end{document}